\pdfoutput=1
\documentclass[12pt,a4paper]{article}

\usepackage{ifthen} 
\newboolean{pdflatex}
\setboolean{pdflatex}{true} 

\newboolean{articletitles}
\setboolean{articletitles}{true} 

\newboolean{uprightparticles}
\setboolean{uprightparticles}{false} 

\def\paperauthors{LHCb collaboration} 
\def\paperasciititle{First observation of a doubly charged tetraquark and its neutral partner} 
\def\papertitle{First observation of a doubly charged tetraquark and its neutral partner} 
\def\paperkeywords{{High Energy Physics}, {LHCb}} 
\def\papercopyright{\the\year\ CERN for the benefit of the LHCb collaboration} 
\def\paperlicence{CC BY 4.0 licence}
\def\paperlicenceurl{https://creativecommons.org/licenses/by/4.0/}

\usepackage[top=1in, bottom=1.25in, left=1in, right=1in]{geometry}

\usepackage{microtype}
\usepackage{lineno}  
\usepackage{xspace} 
\usepackage{caption} 

\usepackage{graphicx}  
\usepackage{color}
\usepackage{colortbl}
\graphicspath{{./figs/}} 

\usepackage{amsmath} 
\usepackage{amssymb}
\usepackage{amsfonts}
\usepackage{upgreek} 

\newcommand*\patchAmsMathEnvironmentForLineno[1]{%
\expandafter\let\csname old#1\expandafter\endcsname\csname #1\endcsname
\expandafter\let\csname oldend#1\expandafter\endcsname\csname
end#1\endcsname
 \renewenvironment{#1}%
   {\linenomath\csname old#1\endcsname}%
   {\csname oldend#1\endcsname\endlinenomath}%
}
\newcommand*\patchBothAmsMathEnvironmentsForLineno[1]{%
  \patchAmsMathEnvironmentForLineno{#1}%
  \patchAmsMathEnvironmentForLineno{#1*}%
}
\AtBeginDocument{%
\patchBothAmsMathEnvironmentsForLineno{equation}%
\patchBothAmsMathEnvironmentsForLineno{align}%
\patchBothAmsMathEnvironmentsForLineno{flalign}%
\patchBothAmsMathEnvironmentsForLineno{alignat}%
\patchBothAmsMathEnvironmentsForLineno{gather}%
\patchBothAmsMathEnvironmentsForLineno{multline}%
\patchBothAmsMathEnvironmentsForLineno{eqnarray}%
}

\usepackage{hyperxmp}

\usepackage[pdftex,
            pdfauthor={\paperauthors},
            pdftitle={\paperasciititle},
            pdfkeywords={\paperkeywords},
            pdfcopyright={Copyright (C) \papercopyright},
            pdflicenseurl={\paperlicenceurl}]{hyperref}
\usepackage[colorinlistoftodos,textsize=scriptsize]{todonotes}

\usepackage[bottom,flushmargin,hang,multiple]{footmisc}

\usepackage[all]{hypcap} 

\usepackage{xspace} 
\usepackage{upgreek}

\def\lhcb   {\mbox{LHCb}\xspace}

\def\MagUp {\mbox{\em Mag\kern -0.05em Up}\xspace}

\ifthenelse{\boolean{uprightparticles}}%
{

 \def\Ppi         {\ensuremath{\uppi}\xspace}

 \def\PDelta      {\ensuremath{\Delta}\xspace}                 
 \def\PXi         {\ensuremath{\Xi}\xspace}                 
 \def\PLambda     {\ensuremath{\Lambda}\xspace}                 
 \def\PSigma      {\ensuremath{\Sigma}\xspace}                 
 \def\POmega      {\ensuremath{\Omega}\xspace}                 
 \def\PUpsilon    {\ensuremath{\Upsilon}\xspace}
 \let\oldPi\Pi
 \def\PPi         {\ensuremath{\oldPi}\xspace}

 \def\PB      {\ensuremath{\mathrm{B}}\xspace}                 
                  
 \def\PD      {\ensuremath{\mathrm{D}}\xspace}

 \def\PK      {\ensuremath{\mathrm{K}}\xspace}

 \def\Pi      {\ensuremath{\mathrm{i}}\xspace}

 \def\Pp      {\ensuremath{\mathrm{p}}\xspace}

 \def\Ps      {\ensuremath{\mathrm{s}}\xspace}

 \def\thebaroffset{0.0em}
}
{

 \def\Ppi         {\ensuremath{\pi}\xspace}

 \mathchardef\PDelta="7101
 \mathchardef\PXi="7104
 \mathchardef\PLambda="7103
 \mathchardef\PSigma="7106
 \mathchardef\POmega="710A
 \mathchardef\PUpsilon="7107
 \mathchardef\PPi="7105
                  
 \def\PB      {\ensuremath{B}\xspace}                 
                  
 \def\PD      {\ensuremath{D}\xspace}

 \def\PK      {\ensuremath{K}\xspace}

 \def\Pi      {\ensuremath{i}\xspace}

 \def\Pp      {\ensuremath{p}\xspace}

 \def\Ps      {\ensuremath{s}\xspace}

 \def\thebaroffset{0.18em}
}
\newcommand{\offsetoverline}[2][\thebaroffset]{\kern #1\overline{\kern -#1 #2}}%

\makeatletter
\ifcase \@ptsize \relax
  \newcommand{\miniscule}{\@setfontsize\miniscule{4}{5}}
\or
  \newcommand{\miniscule}{\@setfontsize\miniscule{5}{6}}
\or
  \newcommand{\miniscule}{\@setfontsize\miniscule{5}{6}}
\fi
\makeatother

\DeclareRobustCommand{\optbar}[1]{\shortstack{{\miniscule (\rule[.5ex]{1.25em}{.18mm})}
  \\ [-.7ex] $#1$}}

\def\squark    {{\ensuremath{\Ps}}\xspace}

\def\pion   {{\ensuremath{\Ppi}}\xspace}

\def\pip    {{\ensuremath{\pion^+}}\xspace}
\def\pim    {{\ensuremath{\pion^-}}\xspace}
\def\pipm   {{\ensuremath{\pion^\pm}}\xspace}

\def\kaon    {{\ensuremath{\PK}}\xspace}

\def\KorKbar {\kern \thebaroffset\optbar{\kern -\thebaroffset \PK}{}\xspace}

\def\Kp      {{\ensuremath{\kaon^+}}\xspace}
\def\Km      {{\ensuremath{\kaon^-}}\xspace}

\def\Dbar    {{\ensuremath{\offsetoverline{\PD}}}\xspace}
\def\D       {{\ensuremath{\PD}}\xspace}

\def\DorDbar {\kern \thebaroffset\optbar{\kern -\thebaroffset \PD}\xspace}
\def\Dz      {{\ensuremath{\D^0}}\xspace}
\def\Dzb     {{\ensuremath{\Dbar{}^0}}\xspace}
\def\Dp      {{\ensuremath{\D^+}}\xspace}
\def\Dm      {{\ensuremath{\D^-}}\xspace}

\def\DpDm    {\ensuremath{\Dp {\kern -0.16em \Dm}}\xspace}

\def\Dstarp  {{\ensuremath{\D^{*+}}}\xspace}
\def\Dstarm  {{\ensuremath{\D^{*-}}}\xspace}

\def\Ds      {{\ensuremath{\D^+_\squark}}\xspace}
\def\Dsp     {{\ensuremath{\D^+_\squark}}\xspace}

\def\B       {{\ensuremath{\PB}}\xspace}

\def\BorBbar {\kern \thebaroffset\optbar{\kern -\thebaroffset \PB}\xspace}
\def\Bz      {{\ensuremath{\B^0}}\xspace}

\def\Bd      {{\ensuremath{\B^0}}\xspace}

\def\BdorBdbar {\kern \thebaroffset\optbar{\kern -\thebaroffset \Bd}\xspace}
\def\Bu      {{\ensuremath{\B^+}}\xspace}
\def\Bub     {{\ensuremath{\B^-}}\xspace}
\def\Bp      {{\ensuremath{\Bu}}\xspace}
\def\Bm      {{\ensuremath{\Bub}}\xspace}

\def\Bs      {{\ensuremath{\B^0_\squark}}\xspace}

\def\BsorBsbar {\kern \thebaroffset\optbar{\kern -\thebaroffset \Bs}\xspace}

\def\Y#1S{\ensuremath{\PUpsilon{(#1S)}}\xspace}

\def\proton      {{\ensuremath{\Pp}}\xspace}

\def\LorLbar     {\kern \thebaroffset\optbar{\kern -\thebaroffset \PLambda}\xspace}

\newcommand{\decay}[2]{\ensuremath{#1\!\to #2}\xspace} 

\def\to                 {\ensuremath{\rightarrow}\xspace}

\def\AT#1     {\ensuremath{A_{\mathrm{T}}^{#1}}\xspace}           

\def\C#1      {\ensuremath{\mathcal{C}_{#1}}\xspace}                       
\def\Cp#1     {\ensuremath{\mathcal{C}_{#1}^{'}}\xspace}                    
\def\Ceff#1   {\ensuremath{\mathcal{C}_{#1}^{\mathrm{(eff)}}}\xspace}        
\def\Cpeff#1  {\ensuremath{\mathcal{C}_{#1}^{'\mathrm{(eff)}}}\xspace}       
\def\Ope#1    {\ensuremath{\mathcal{O}_{#1}}\xspace}                       
\def\Opep#1   {\ensuremath{\mathcal{O}_{#1}^{'}}\xspace}                    

\newcommand{\aunit}[1]{\ensuremath{\text{\,#1}}}       

\newcommand{\tev}{\aunit{Te\kern -0.1em V}\xspace}
\newcommand{\gev}{\aunit{Ge\kern -0.1em V}\xspace}
\newcommand{\mev}{\aunit{Me\kern -0.1em V}\xspace}
\newcommand{\kev}{\aunit{ke\kern -0.1em V}\xspace}
\newcommand{\ev}{\aunit{e\kern -0.1em V}\xspace}
 
\newcommand{\mevc}{\ensuremath{\aunit{Me\kern -0.1em V\!/}c}\xspace}
\newcommand{\gevc}{\ensuremath{\aunit{Ge\kern -0.1em V\!/}c}\xspace}
\newcommand{\mevcc}{\ensuremath{\aunit{Me\kern -0.1em V\!/}c^2}\xspace}
\newcommand{\gevcc}{\ensuremath{\aunit{Ge\kern -0.1em V\!/}c^2}\xspace}

\newcommand{\chisq}{\ensuremath{\chi^2}\xspace}

\def\gsim{{~\raise.15em\hbox{$>$}\kern-.85em
          \lower.35em\hbox{$\sim$}~}\xspace}
\def\lsim{{~\raise.15em\hbox{$<$}\kern-.85em
          \lower.35em\hbox{$\sim$}~}\xspace}

\def\evtgen     {\mbox{\textsc{EvtGen}}\xspace}

\def\geant      {\mbox{\textsc{Geant4}}\xspace}

\def\pythia     {\mbox{\textsc{Pythia}}\xspace}

\def\tell1  {TELL1\xspace}
\def\ukl1   {UKL1\xspace}

\newcommand{\lhcborcid}[1]{\href{https://orcid.org/#1}{\hspace*{0.1em}\raisebox{-0.45ex}{\includegraphics[width=1em]{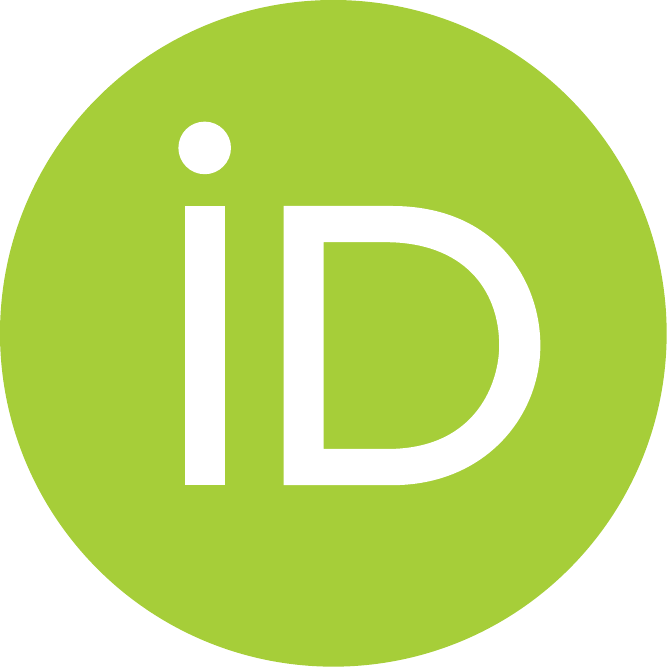}}}}

\usepackage{cite} 
\usepackage{mciteplus}
\usepackage{longtable} 

\begin{document}

\renewcommand{\thefootnote}{\fnsymbol{footnote}}
\setcounter{footnote}{1}

\begin{titlepage}
\pagenumbering{roman}

\vspace*{-1.5cm}
\centerline{\large EUROPEAN ORGANIZATION FOR NUCLEAR RESEARCH (CERN)}
\vspace*{1.5cm}
\noindent
\begin{tabular*}{\linewidth}{lc@{\extracolsep{\fill}}r@{\extracolsep{0pt}}}
\ifthenelse{\boolean{pdflatex}}
{\vspace*{-1.5cm}\mbox{\!\!\!\includegraphics[width=.14\textwidth]{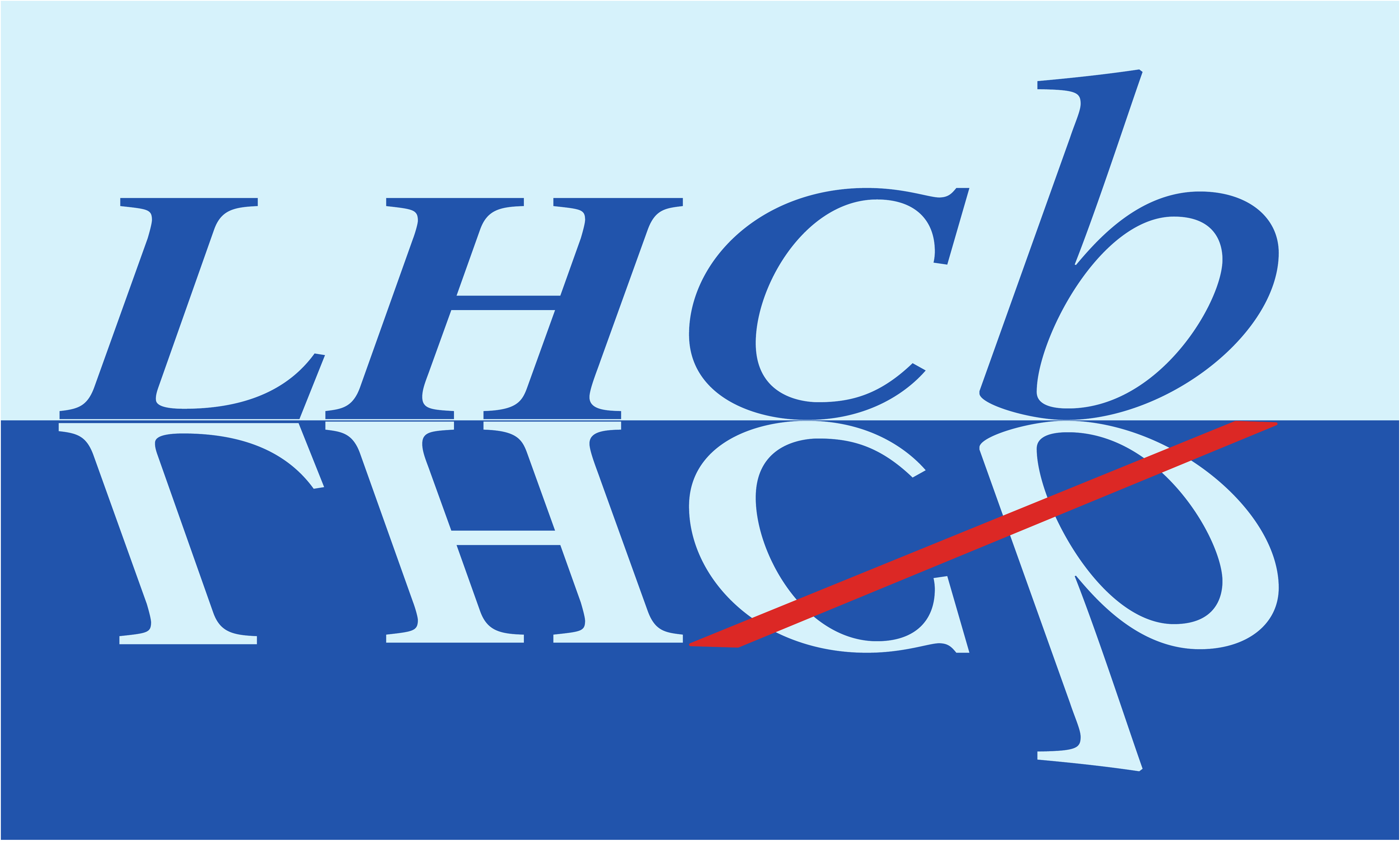}} & &}%
{\vspace*{-1.2cm}\mbox{\!\!\!\includegraphics[width=.12\textwidth]{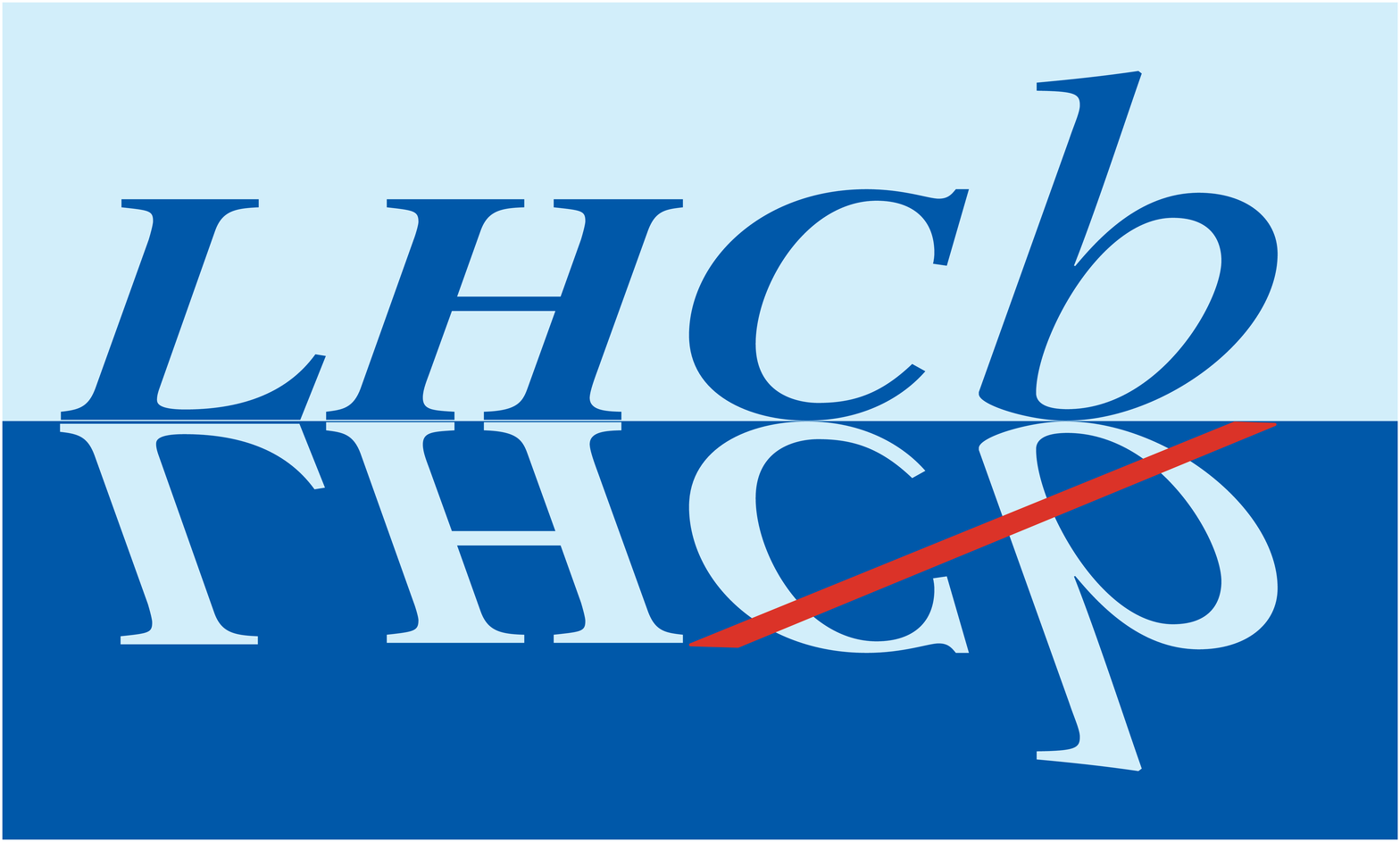}} & &}%
\\
 & & CERN-EP-2022-239 \\  
 & & LHCb-PAPER-2022-026 \\  
 & & July 27, 2023 \\ 
 & & \\
\end{tabular*}

\vspace*{4.0cm}

{\normalfont\bfseries\boldmath\huge
\begin{center}
  \papertitle 
\end{center}
}

\vspace*{2.0cm}

\begin{center}
\paperauthors\footnote{Authors are listed at the end of this letter.}
\end{center}

\vspace{\fill}

\begin{abstract}
  \noindent
A combined amplitude analysis is performed for the decays $B^0 \rightarrow \offsetoverline{D}{}^0 D^+_s\pi^-$ and $B^+\rightarrow D^- D^+_s\pi^+$, which are related by isospin symmetry. The analysis is based on data collected
by the LHCb detector in proton-proton collisions at center-of-mass energies of 7, 8 and 13 TeV. The full data sample corresponds to an integrated luminosity of 9\,$\rm{fb^{-1}}$. 
Two new resonant states with masses of $2.908\pm0.011\pm0.020\,\rm{GeV}$ and widths of $0.136\pm0.023\pm0.013\,\rm{GeV}$ are observed, which decay to $D^+_s\pi^+$ and $D^+_s\pi^-$ respectively.
The former state indicates the first observation of a doubly charged open-charm tetraquark state with minimal quark content $[c\bar{s}u\bar{d}]$, and the latter state is a neutral tetraquark composed of $[c\bar{s}\bar{u}d]$ quarks. Both states are found to have spin-parity $0^+$,
and their resonant parameters are consistent with each other, which suggests that they belong to an isospin triplet.

\end{abstract}

\vspace*{2.0cm}

\begin{center}
  Published in \href{https://journals.aps.org/prl/abstract/10.1103/PhysRevLett.131.041902}{Phys. Rev. Lett. 131 (2023) 041902}
\end{center}

\vspace{\fill}

{\footnotesize 
\centerline{\copyright~\papercopyright. \href{\paperlicenceurl}{\paperlicence}.}}
\vspace*{2mm}

\end{titlepage}


\newpage
\setcounter{page}{2}
\mbox{~}

\renewcommand{\thefootnote}{\arabic{footnote}}
\setcounter{footnote}{0}

\cleardoublepage


\pagestyle{plain} 
\setcounter{page}{1}
\pagenumbering{arabic}


\def\Zz         {{\ensuremath{T^a_{c\bar{s}0}(2900)}}\xspace}
\def\Zzz        {{\ensuremath{T^a_{c\bar{s}0}(2900)^0}}\xspace}
\def\Zzpp       {{\ensuremath{T^a_{c\bar{s}0}(2900)^{++}}}\xspace}

\def\DzKpi      {{\ensuremath{\D^0_{K\pi}}}\xspace}
\def\DzbKpi     {{\ensuremath{\Dbar{}^0_{K\pi}}}\xspace}
\def\DzKtpi      {{\ensuremath{\D^0_{K3\pi}}}\xspace}
\def\DzbKtpi     {{\ensuremath{\Dbar{}^0_{K3\pi}}}\xspace}

\def\BztoDzbarDsppim        {{\ensuremath{\Bz \rightarrow \Dzb \Dsp \pim}}\xspace}
\def\BztoDzbarKpiDsppim     {{\ensuremath{\Bz \rightarrow \DzbKpi \Dsp \pim}}\xspace}
\def\BztoDzbarKtpiDsppim    {{\ensuremath{\Bz \rightarrow \DzbKtpi \Dsp \pim}}\xspace}
\def\BptoDmDsppip           {{\ensuremath{\Bp \rightarrow \Dm \Dsp \pip}}\xspace}

In the quark model, tetraquark hadrons are predicted in addition to the conventional mesons and baryons~\cite{Gell-Mann:1964ewy}.
States composed of four different quark types present unambiguous evidence for tetraquarks~\cite{Molina:2010tx} and have attracted great interest in studies of hadron spectroscopy~\cite{Zhu:2022arXiv}.
The discoveries of the $D^{*}_{s0}(2317)^+$~\cite{BaBar:2003oey} and $D_{s1}(2460)^+$~\cite{CLEO:2003ggt} states prompted speculation that they may have a tetraquark component~\cite{PDG2022,Chen:2016spr}, but no evidence for isospin partners was found in explicit searches~\cite{BaBar:2006eep,Belle:2015glz}. The D0 collaboration claimed evidence for an $X(5568)$ state with quark configuration of $[b\bar{s}u\bar{d}]$ in decays to $B_s^0\pi^\pm$~\cite{D0:2016mwd,D0:2017qqm}.
Its existence, however, has not been confirmed by other experiments~\cite{LHCb-PAPER-2016-029,CMS:2017hfy,ATLAS:2018udc,CDF:2017dwr}. 

In the open charm sector, it is natural to search for $D_s^+\pi^\pm$ resonances, which would be candidate tetraquark states with minimal quark content of $[c\bar{s}u\bar{d}]$ and $[c\bar{s}\bar{u}d]$
as predicted in the diquark-antidiquark model~\cite{Agaev:2016lkl,Chen:2017rhl}. 
In particular, an attractive potential in the $\Dsp\pi^+$ channel facilitates searches for an open-charm tetraquark state according to lattice QCD calculation~\cite{Baeza-Ballesteros:2022azb}.

In 2020, two new resonant structures, $X_0(2900)$ and $X_1(2900)$, were observed by the LHCb collaboration in the $D^-K^+$ mass spectrum of the $\decay{B^+}{D^+D^-K^+}$ decay~\cite{LHCb-PAPER-2020-024,LHCb-PAPER-2020-025}, motivating theoretical speculation as to their nature~\cite{Zhu:2022arXiv}. 
Predictions of a doubly charged tetraquark $[c\bar{s}u\bar{d}]$ and its isospin partner $[c\bar{s}\bar{u}d]$ were made based on those observations~\cite{He:2020jna,Lu:2020qmp,Burns:2020xne,Agaev:2021knl,Agaev:2021jsz,Azizi:2021aib}.

The decays $\BztoDzbarDsppim$ and $\BptoDmDsppip$  are ideal channels in which to search for possible exotic states decaying to $D_s\pi$.
The only resonances expected to contribute to the two decays are excited $\Dbar{}^{*}$ resonances decaying to $\Dzb\pim$ and $\Dm\pip$ final states, which have been extensively studied~\cite{PDG2022,Chen:2016spr,LHCB-PAPER-2013-026,LHCB-PAPER-2014-070,LHCB-PAPER-2016-026}. Any resonant structure that cannot be described by the known $\Dbar{}^*$ excited states ($\Dzb\pim$ or $\Dm\pip$) implies the existence of potential exotic objects decaying into either $D_s\pi$ or $DD_s$ final states.\footnote{Charge conjugation is implied throughout this Letter.}

A combined amplitude analysis of the $\BztoDzbarDsppim$ and $\BptoDmDsppip$ decays is performed, where the amplitudes in the two decay modes are related through isospin symmetry.
The analysis is based on proton-proton collision data collected with the \lhcb detector, corresponding to a total integrated luminosity of $9~\rm fb^{-1}$ at center-of-mass energies $7,~8$ and $13\tev$.

The LHCb detector is a single-arm forward spectrometer covering the pseudorapidity range $2 < \eta < 5$, described in detail in Refs.~\cite{LHCb-DP-2008-001,LHCb-DP-2014-002}. It is designed specifically for the study of particles containing $b$ or $c$ quarks. Simulation is required to model the effects of the detector acceptance and the imposed selection requirements. The simulated samples are generated with \pythia~\cite{Sjostrand:2007gs,*Sjostrand:2006za}, \evtgen~\cite{Lange:2001uf} and the \geant toolkits~\cite{Allison:2006ve, *Agostinelli:2002hh} as described in Ref.~\cite{LHCb-PROC-2011-006}.

A brief summary of the selections is given here and further details can be found in Ref.~\cite{PRD_twin}. 
Charmed mesons are reconstructed using the $\Dzb\to\Kp\pim$, $\Dzb\to\Kp\pim\pim\pip$, $\Dm\to\Kp\pim\pim$ and $\Dsp\to\Kp\Km\pip$ decays. The invariant mass $M$ of the $\Dzb\pim$ pair is required to be larger than $2050\mev$ (natural units with $\hbar=c=1$ are used throughout) to veto the majority of the $D^*(2010)^-$ contribution from the $\Bz\to\Dstarm\Dsp$ decay. A multivariate classifier, based on a boosted decision tree (BDT)~\cite{Breiman,AdaBoost} algorithm in the TMVA toolkit~\cite{Hocker:2007ht,*TMVA4}, is employed to reduce combinatorial backgrounds. 
In addition, non-double-charm backgrounds are reduced by requirements on the \D meson mass and by requiring a significant flight distance of the reconstructed \D candidates from the \B decay vertex.

\begin{figure}[!tb]
  \begin{center}
    \includegraphics[width=0.48\linewidth]{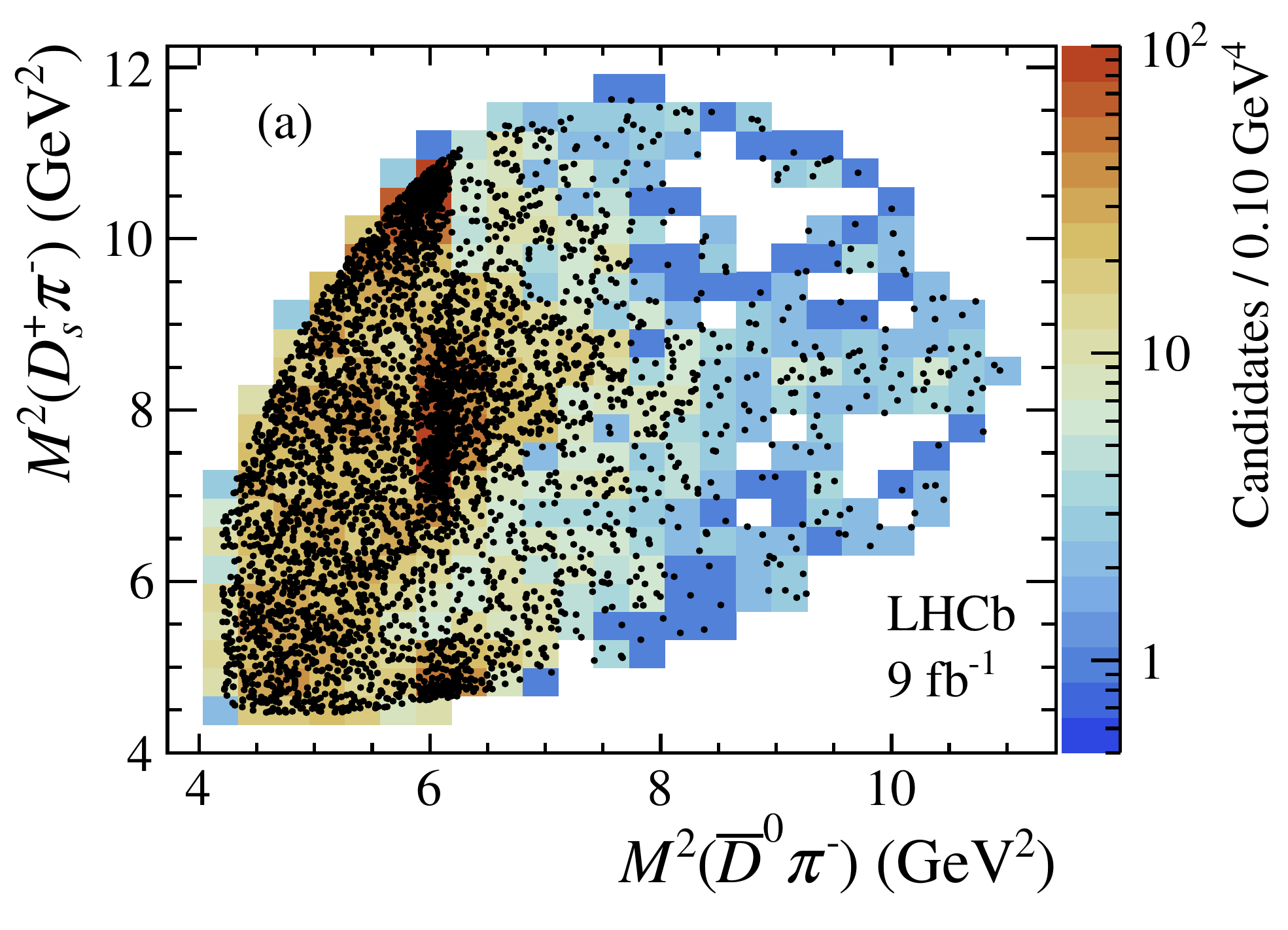}
    \includegraphics[width=0.48\linewidth]{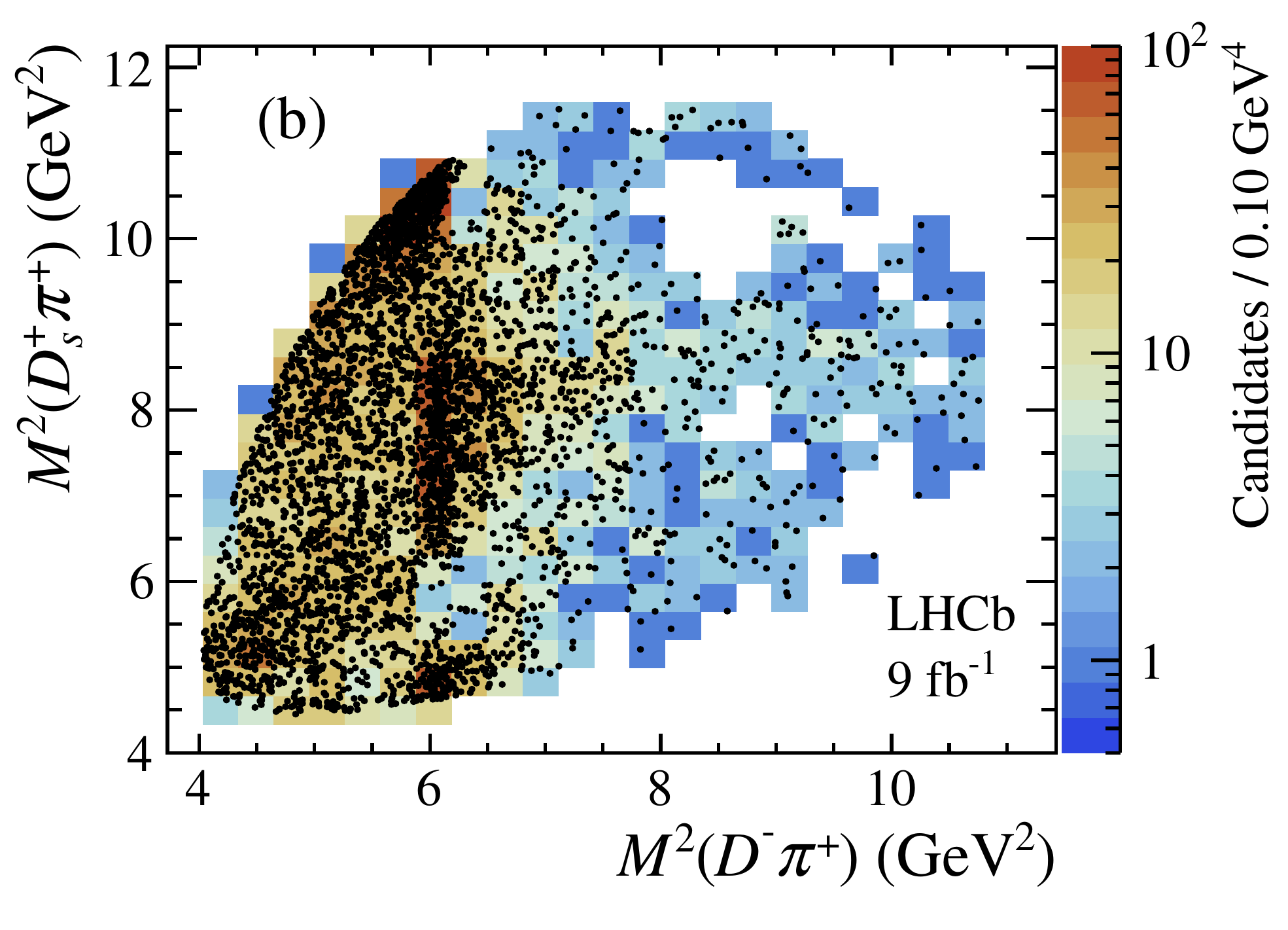}
  \end{center}
  \caption{Dalitz plots of (a) $\decay{\Bz}{\Dzb\Dsp\pim}$ and (b) $\decay{\Bp}{\Dm\Dsp\pip}$ decays after the full selection is applied.}
  \label{fig:Dalitz plots}
\end{figure}

After the above selections, 4009 $\BztoDzbarDsppim$ candidates and 3750 $\BptoDmDsppip$ candidates are retained. Extended maximum likelihood fits are performed to the $\Dzb \Dsp \pim$ and $\Dm \Dsp \pip$ invariant-mass distributions separately in order to extract signal yields, all in the range $[5230, 5630]\mev$. 
The purities of the $\BztoDzbarDsppim$ and $\BptoDmDsppip$ samples are $90.7\%$ and $95.2\%$, respectively. Dalitz plots for the two decays, shown in Fig.~\ref{fig:Dalitz plots}, show similar features as a result of the isospin symmetry of the decays, indicating the feasibility of a combined treatment. The vertical band at $M^2(\Dbar\pi)\approx 6\gev^2$ corresponds to the $J^P=2^+$ state $D_2^*(2460)$ ($D_2^*(2460)^-$ in $\BztoDzbarDsppim$ and $\Dbar{}_2^*(2460)^0$ in $\BptoDmDsppip$ decays). Furthermore, a faint horizontal band at $M^2(\Ds\pi)\approx 8.5\gev^2$ is visible in both plots.

The conventional contributions anticipated in the two decays include tails from the $D^*(2010)^-$ and $\Dbar{}^*(2007)^0$ resonances in $\BptoDmDsppip$ and $\BztoDzbarDsppim$ decays, respectively, in addition to the higher excited $\Dbar{}_2^*(2460)$, $\Dbar{}_1^*(2600)$, $\Dbar{}_3^*(2750)$, $\Dbar{}_1^*(2760)$ 
and $\Dbar(3000)$ states~\cite{PDG2022}.
In previous measurements, only the neutral $\Dbar{}_1^*(2600)^0$, $\Dbar{}_1^*(2760)^0$ and $\Dbar(3000)^0$ mesons were observed, while their charged partners were not. 
For the $\Dbar\pi$ S-wave components, a quasi-model-independent (QMI) method~\cite{LHCB-PAPER-2016-026} 
is applied where 11 spline points are chosen in the $M(\Dbar\pi)$ spectrum, at
$(1.9, 2.0, 2.1, 2.2, 2.3, 2.4, 2.5, 2.6, 2.7, 2.9, 3.4)\gev$.
As isospin symmetry is imposed, these excited $\Dbar{}^{*}$ states are added in both channels before testing the exotic states decaying either to the $\Ds\pi$ or $\Dbar\Ds$ final state. 

The complex amplitude $\mathcal{A}_i(x;\Theta_i)$ for the decay through each intermediate state $i$ is constructed using the helicity formalism, where $x$ denotes variables calculated from the four-momenta of the final-state particles in each event, and $\Theta$ is a set of model parameters. A relativistic Breit--Wigner function is used to describe the lineshape of the resonant states, where the resonant parameters are fixed to their known values~\cite{PDG2022}. The total amplitude $\mathcal{A}(x;\Theta)$ is the sum of the complex amplitudes for each contribution. The spin-parity of the $\Dbar(3000)^0$ state is currently unknown. The $4^+$ spin-parity assignment is found to be the most likely~\cite{PRD_twin} and is used in the default model here.
In the QMI component, the amplitudes of the first and last spline points are fixed to 0.

A negative-log-likelihood, $-\sum_{j}^{N} \ln P\left(x_{j};\Theta\right)$, 
is minimized to determine the fit parameters, where $N$ denotes the number of $B$ meson candidates. The probability density function, PDF, is defined as
\begin{equation}
    P\left(x;\Theta\right) = f_{\rm sig}\cdot P^{\rm norm}_{\rm sig}(x;\Theta) + f_{\rm bkg}\cdot P^{\rm norm}_{\rm bkg}(x),
\end{equation}
where the fractions are determined from the fit to the invariant mass distribution of \B candidates. The normalized PDF for the signal is expressed as
\begin{equation}
    P^{\rm norm}_{\rm sig}(x;\Theta) = \frac{\epsilon(x)\vert\mathcal{A}\left(x;\Theta\right)\vert^{2}}{I_{\rm sig}\left(\Theta\right)}.
\end{equation}
The factor $I_{\rm sig}(\Theta)$ is the signal PDF normalisation, obtained using a simulated sample generated uniformly over the phase space. The efficiency map, $\epsilon(x)$, is the signal efficiency derived from the simulated samples after applying all the selection criteria, and smoothed using a kernel density estimation method~\cite{Poluektov:2014rxa}. 
The background PDF is 
obtained from the Dalitz plot distribution of data in the ranges
$[5400, 5700]\mev$ for \Bp decays, and $[5500, 5700]\mev$ for \Bz decays to avoid tails from $\Bs \rightarrow \Dzb \Dsp \pim$ decays.
The extrapolation into the signal region of $|M(\B) - m(\B)|<20\mev$ is based on a Gaussian process~\cite{mathad2021efficient}. Here $M(\B)$ and $m(\B)$ are the candidate mass and known mass~\cite{PDG2022} of \Bz or \Bp mesons. 

\begin{figure}[!tb]
  \begin{center}
    \includegraphics[width=0.33\linewidth]{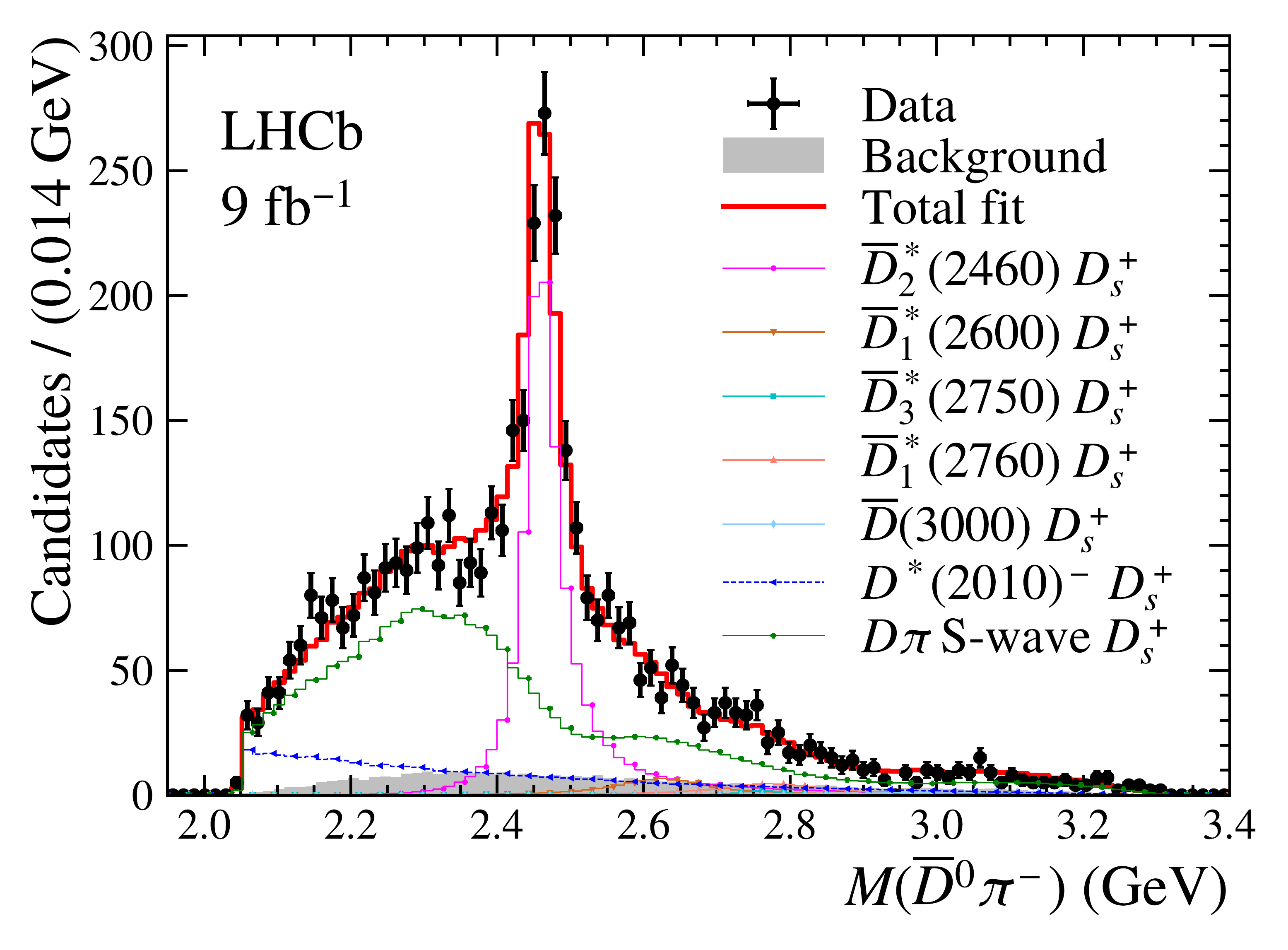}\put(-105,90){(a)}
    \includegraphics[width=0.33\linewidth]{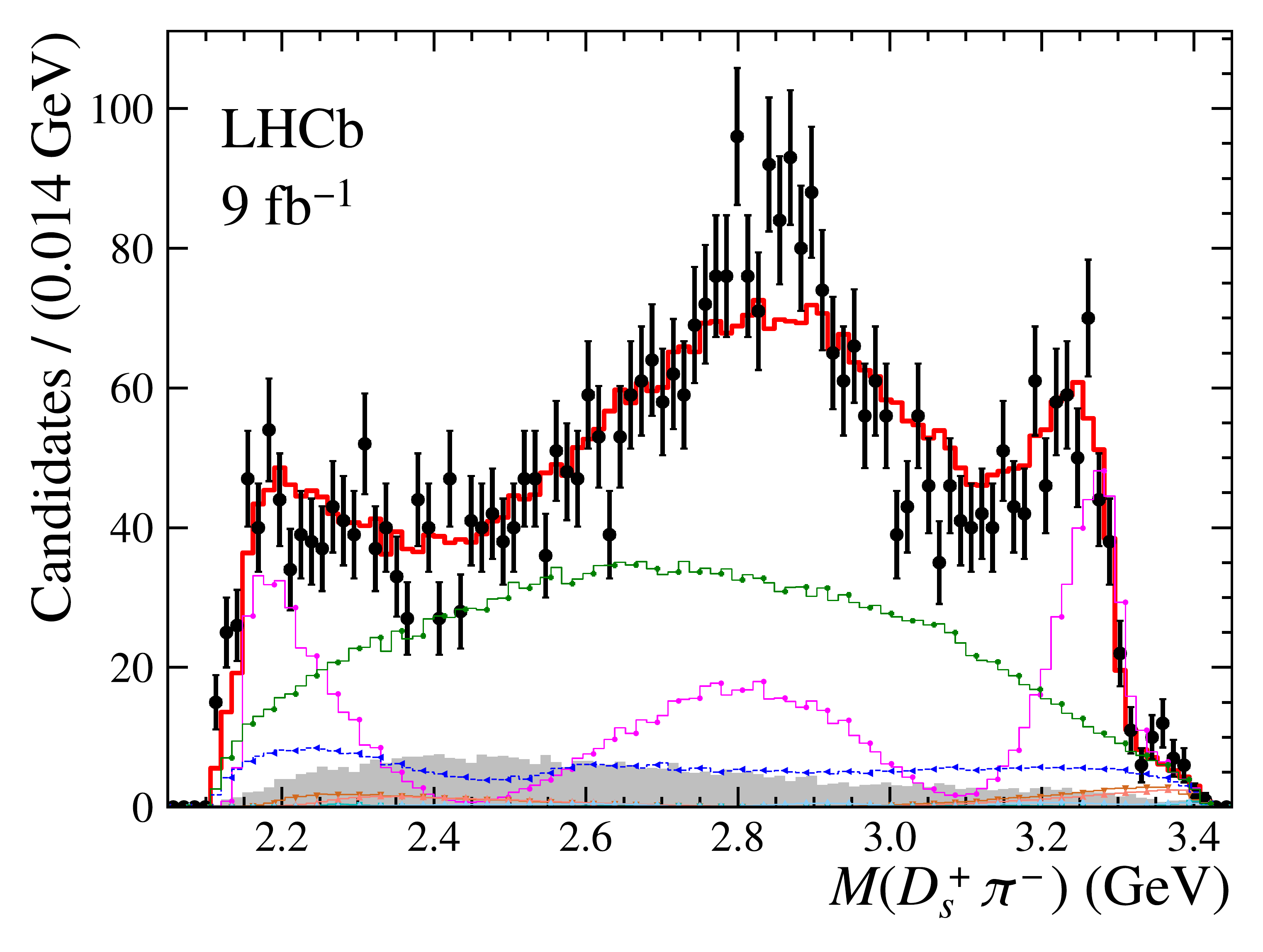}\put(-105,90){(b)}
    \includegraphics[width=0.33\linewidth]{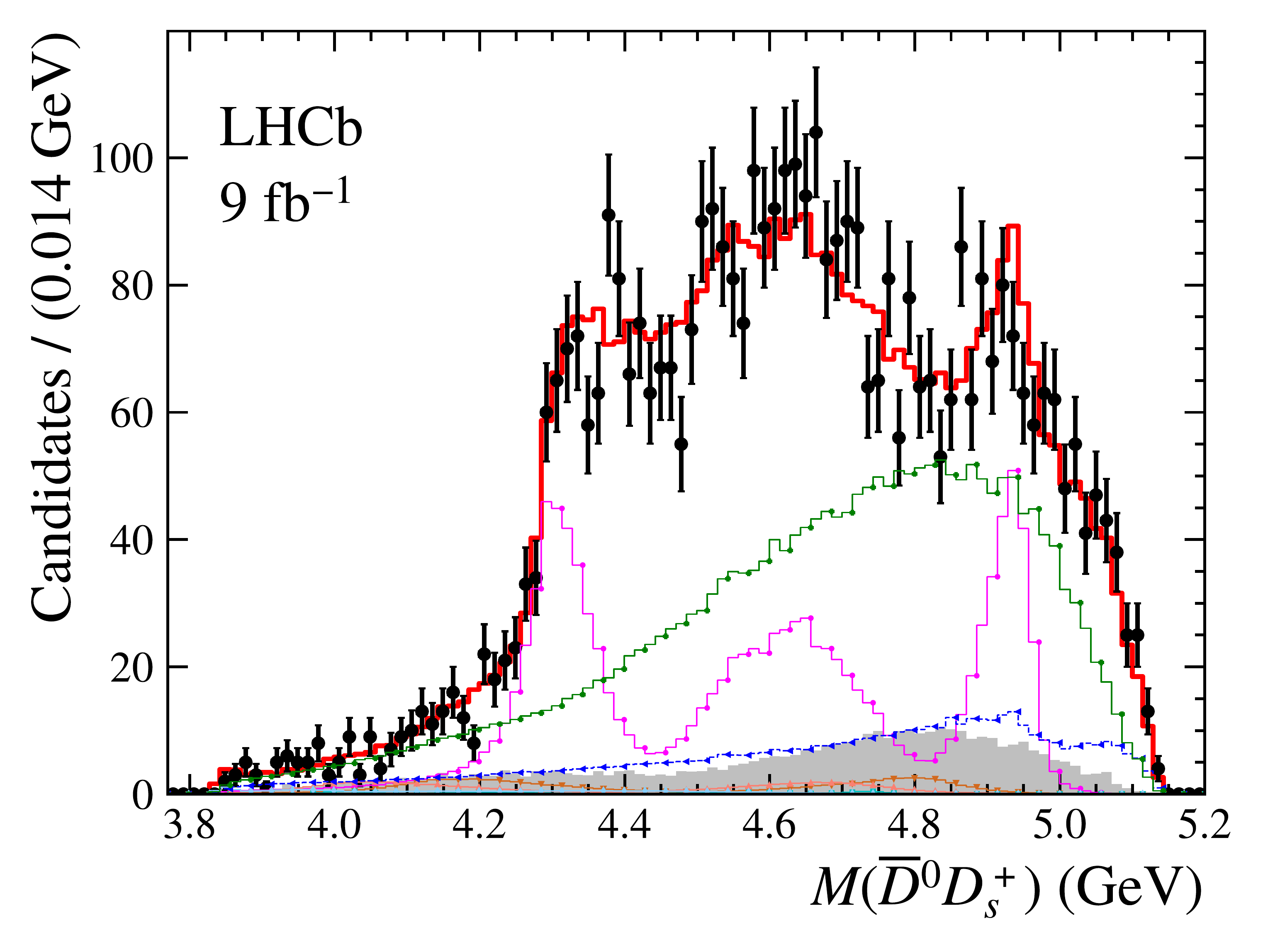}\put(-105,90){(c)}\\
    \includegraphics[width=0.33\linewidth]{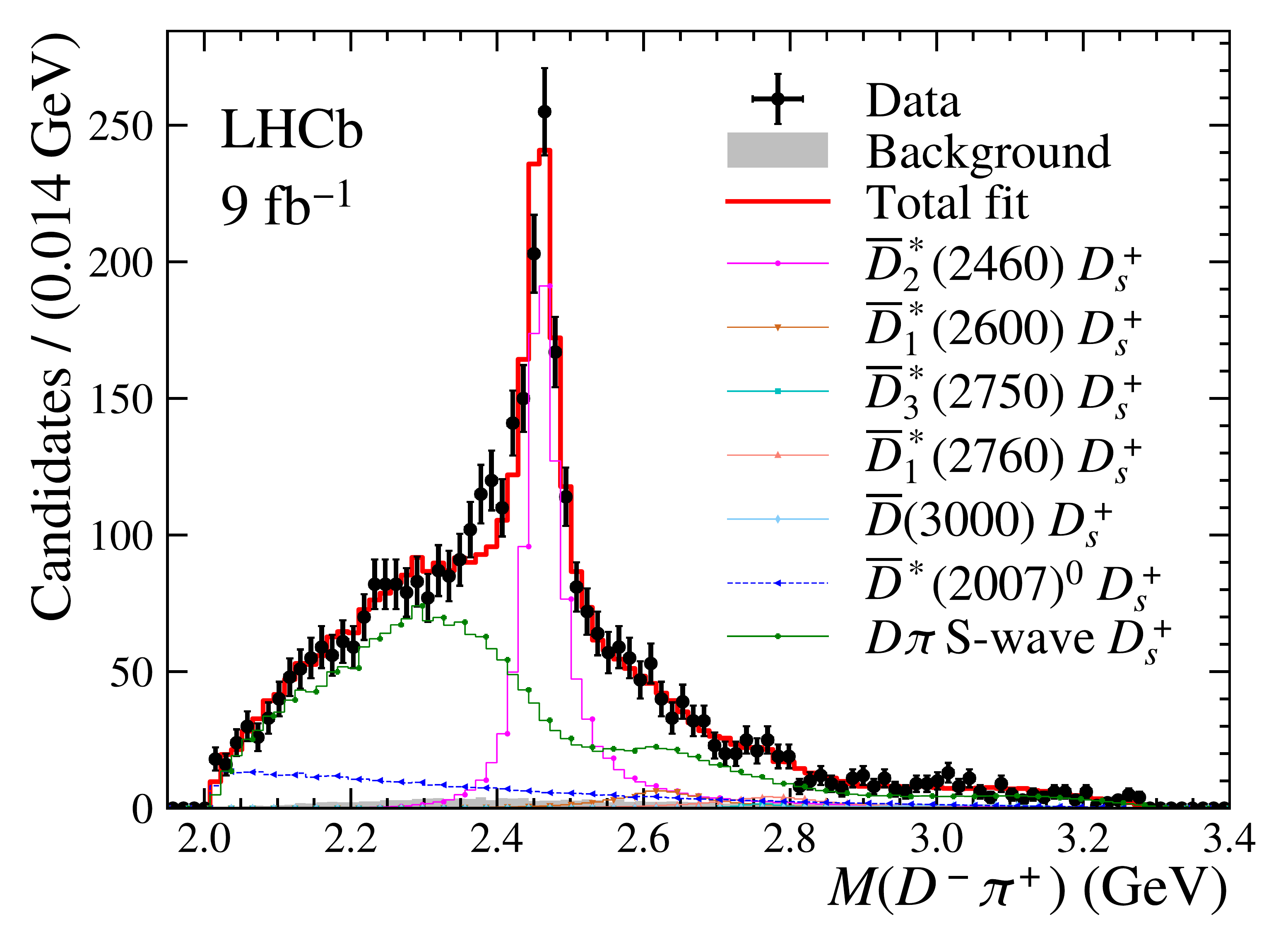}\put(-105,90){(d)}
    \includegraphics[width=0.33\linewidth]{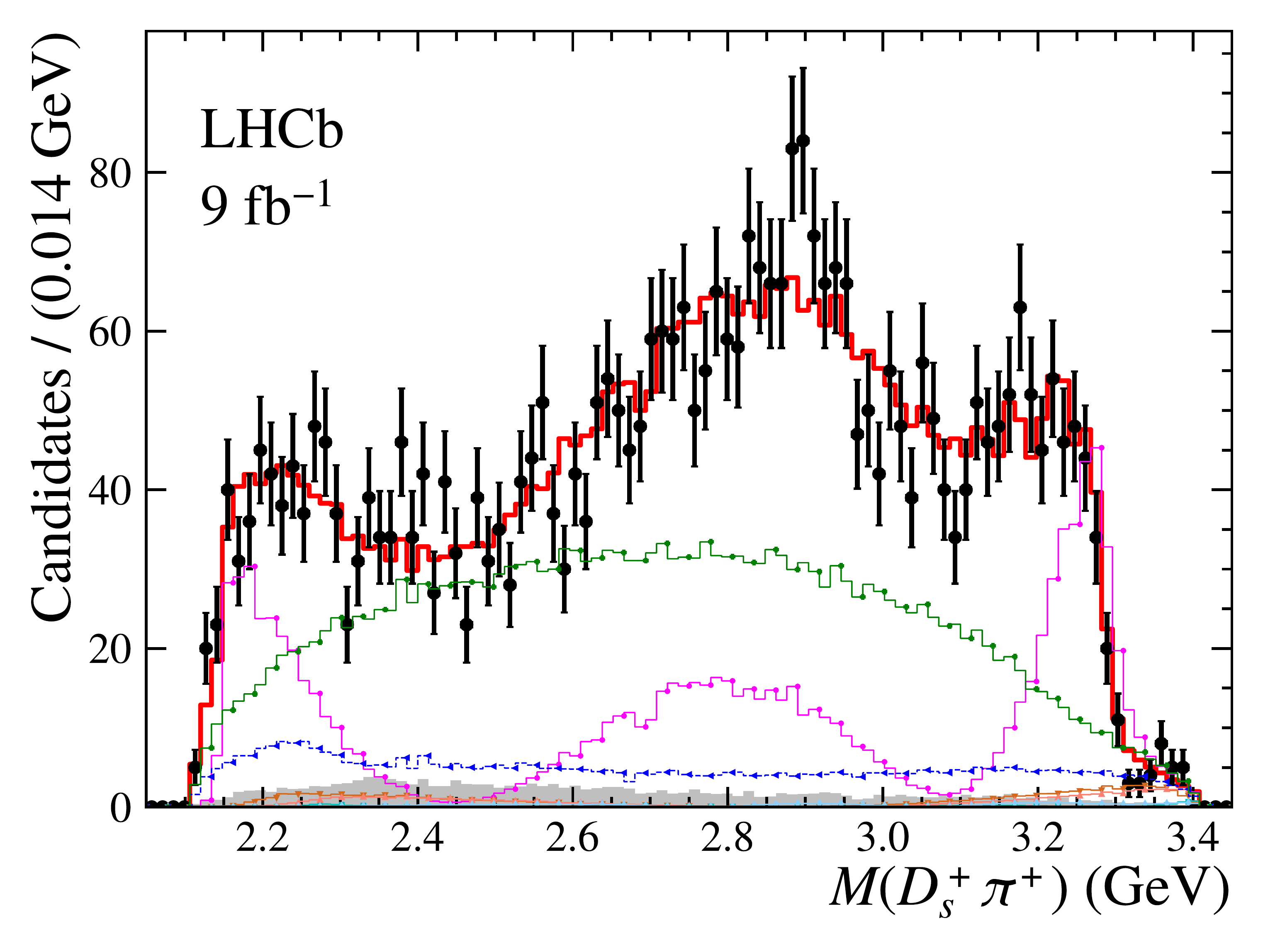}\put(-105,90){(e)}
    \includegraphics[width=0.33\linewidth]{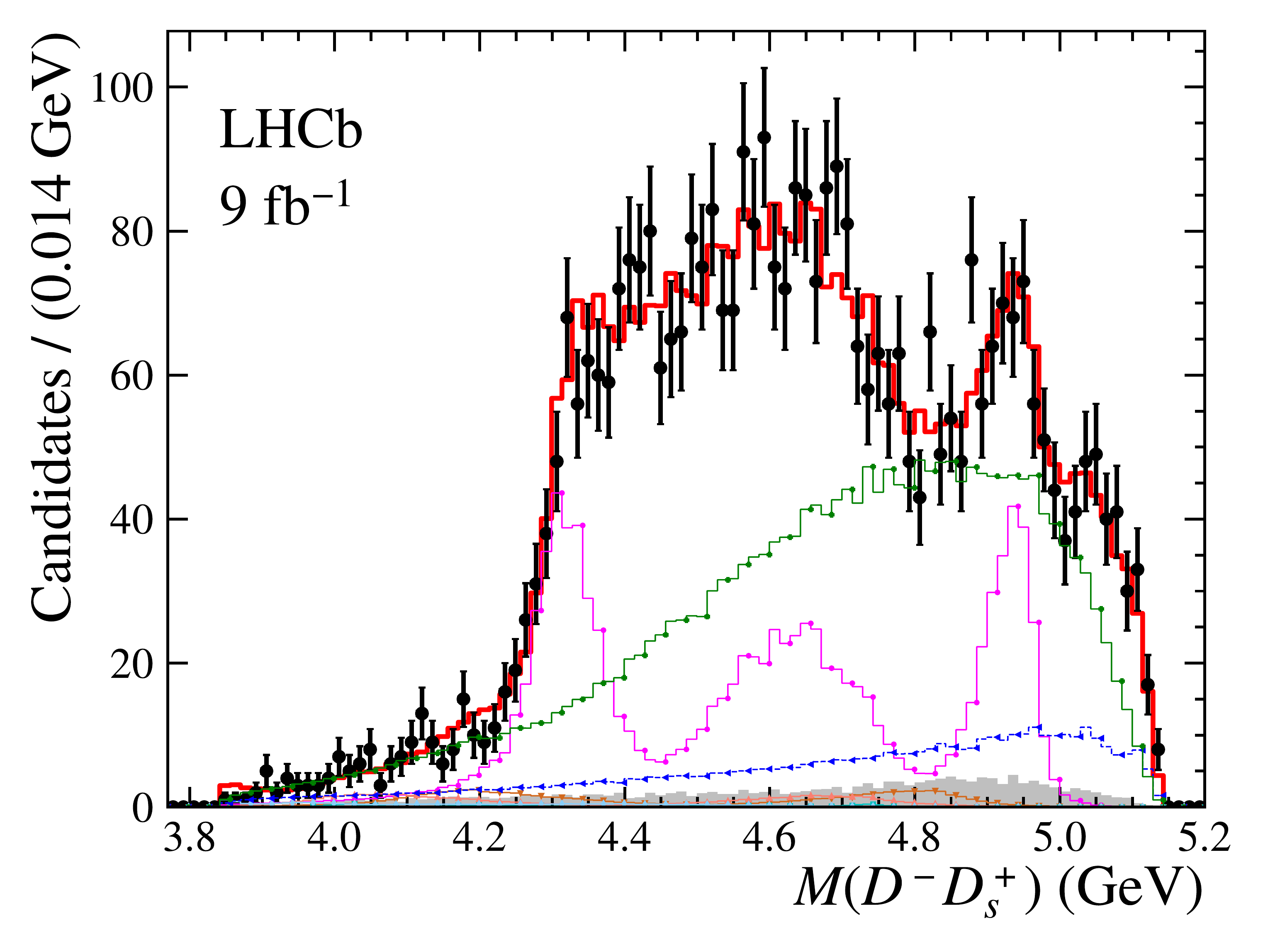}\put(-105,90){(f)}
  \end{center}
  \caption{Distributions of (a) $M(\Dzb\pim)$; (b) $M(\Dsp\pim)$; (c) $M(\Dzb\Dsp)$ for the $\Dbar\pi$ components of the $\BztoDzbarDsppim$ candidates; and (d) $M(\Dm\pip)$; (e) $M(\Dsp\pip)$; (f) $M(\Dm\Dsp)$ of the $\BptoDmDsppip$ candidates. The data are overlaid with the results of the fits.}
  \label{fig:baseline_fit}
\end{figure}

A simultaneous fit is performed to the two decay samples, where all parameters are shared except the masses, widths and complex parameters of different excited $\Dbar{}^*(2007)^0$ and $D^*(2010)^-$ states. 
The projections of the fit results are shown in Fig.~\ref{fig:baseline_fit},  only taking $\Dbar\pi$ components into account. The models describe the data well in the different invariant mass projections, except for the $M(\Ds\pi)$ distributions,
where peaking structures near $2.9\gev$ in the data cannot be attributed to any $\Dbar\pi$ component.
Additional fits are attempted where new $\Dbar{}^*$ states with different spin-parities are added, allowing their masses and widths to vary freely, but no satisfactory description of this region is found. 

Two $\Ds\pi$ states are introduced and, under isospin symmetry, they share the following resonance parameters: the complex amplitude factor, the mass and the width of the states. 
The $M(\Ds\pi)$ distributions of the fit results are shown in Fig.~\ref{fig:Z0_fit_MDspi}, while the other projections are shown in Fig. S1 of the Supplemental Material~\cite{MySupp}. Figure S2 of the Supplemental Material~\cite{MySupp} shows the fit projection onto $M(\Ds\pi)$ with and without the additional $\Ds\pi$ states in the region of $M(\Dbar\pi)>2.7\gev$, where most of the $D^{**}$ contributions are suppressed. 
The peaks in the $M(\Ds\pi)$ distribution near $2.9\gev$, as well as the dip near $3.0\gev$, are better described. 

Different spin-parity scenarios are tested. The result with $J^P =0^+$ $\Ds\pi$ states produces the best likelihood, and is taken as the default fit result. 
The mass and width are determined to be $2.909\pm0.010\gev$ and $0.134\pm0.019\gev$, respectively. The other  parameters of the result are given in Table S1 of the Supplemental Material~\cite{MySupp}.
Following the convention in Ref.~\cite{Gershon:2022xnn}, the new states are named $\Zzz$ and $\Zzpp$. 
The $M(D_s\pi)$ mass resolution is estimated to be approximately $4\mev$ near the $\Zz$ measured mass, which is much smaller than the width of the $\Zz$ state, and is therefore neglected.
\begin{figure}[!tb]
  \begin{center}
    \includegraphics[width=0.40\linewidth]{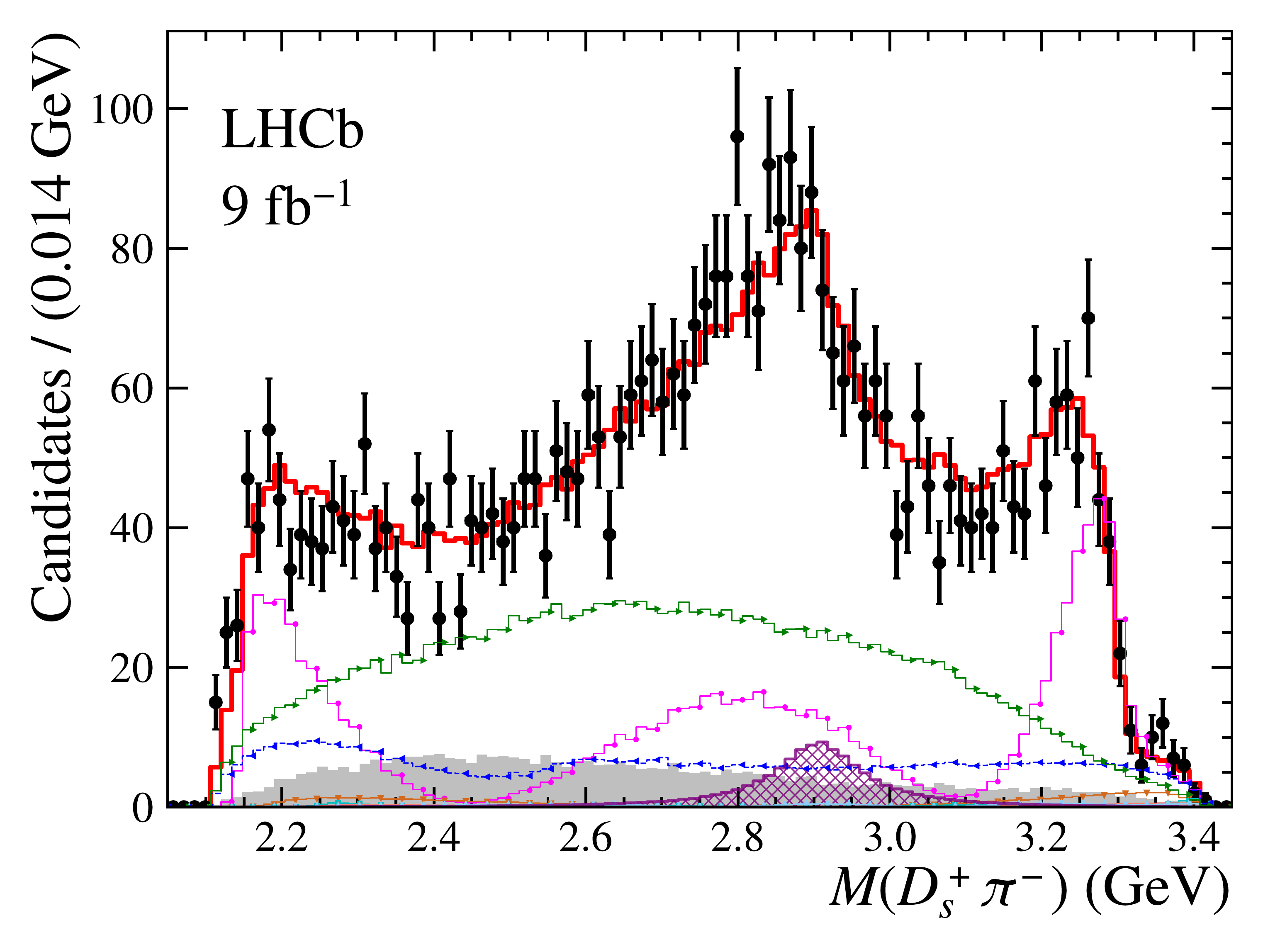}\put(-125,115){(a)}
    \includegraphics[width=0.40\linewidth]{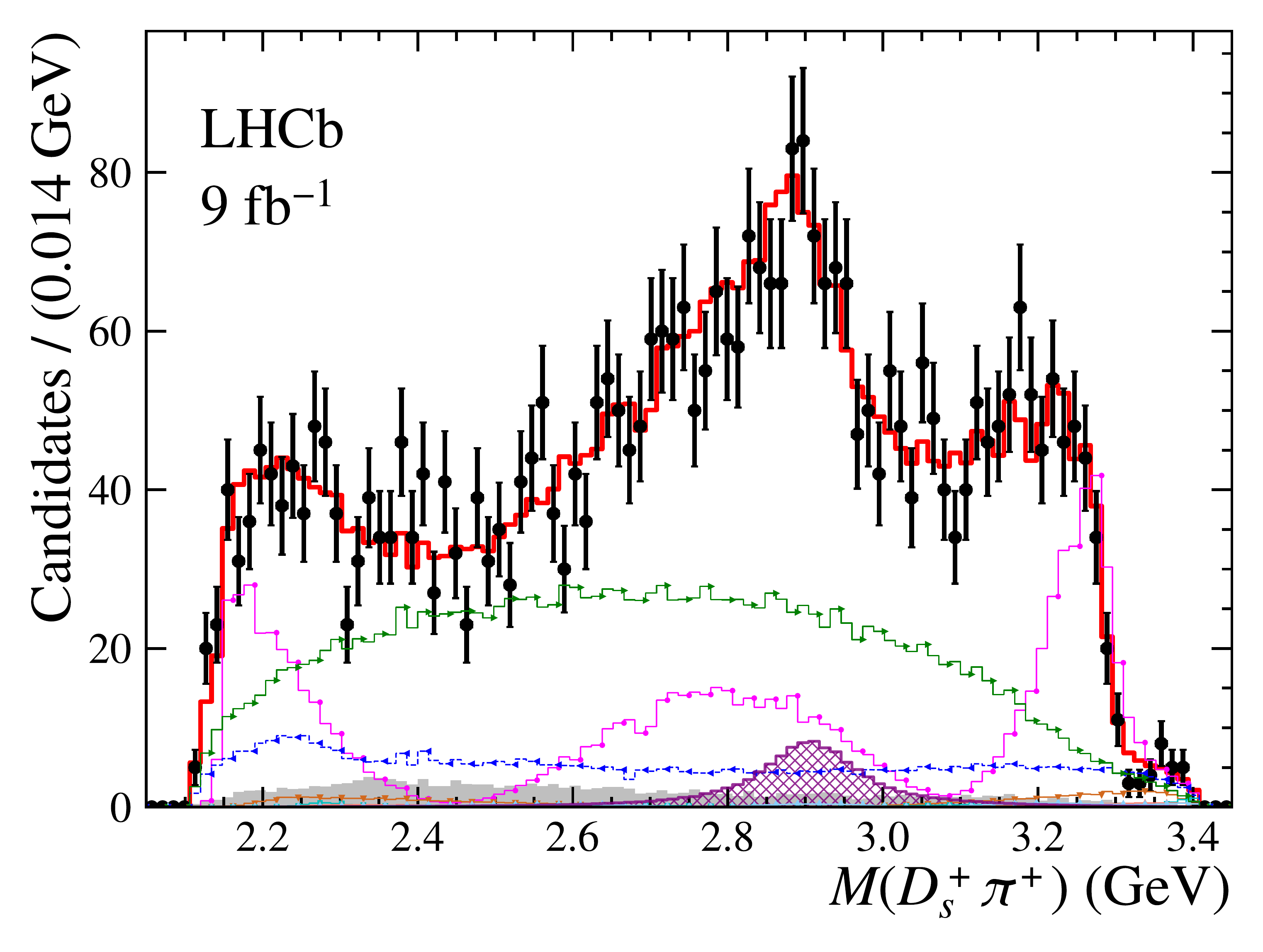}\put(-125,115){(b)}
    \includegraphics[width=0.18\linewidth]{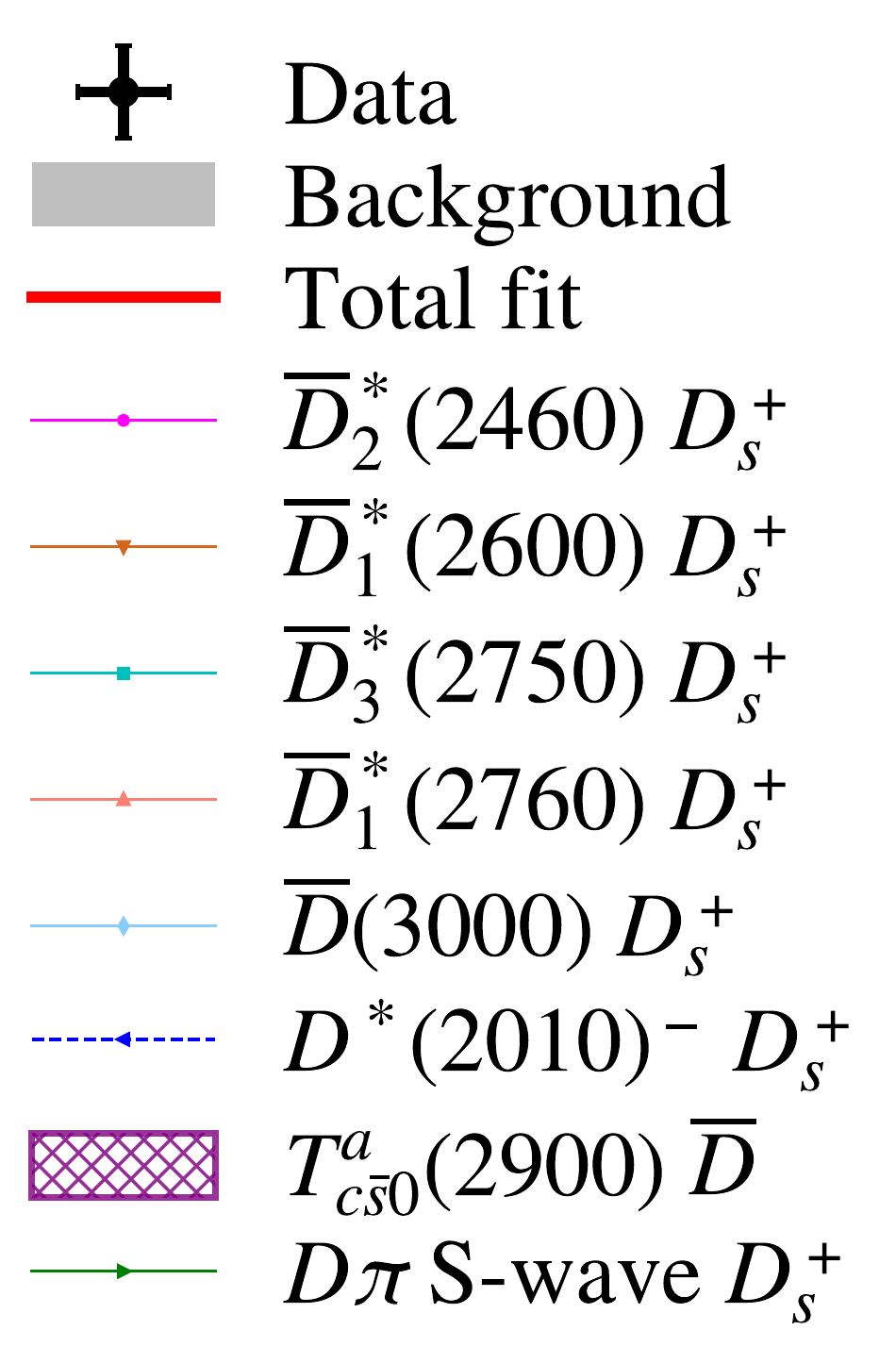}
  \end{center}
  \caption{Distributions of (a) $M(\Dsp\pim)$ of $\BztoDzbarDsppim$ decays; and (b) $M(\Dsp\pip)$ for the $\BptoDmDsppip$ sample. The data are overlaid with the fit results with the inclusion of the new $0^+$ $\Ds\pi$ resonant states.}
  \label{fig:Z0_fit_MDspi}
\end{figure}
When using separate parameters for the $\Zzz$ state in  $\BztoDzbarDsppim$ decays and the $\Zzpp$ state in $\BptoDmDsppip$ decays, without changing the treatment of the other states, 
the masses and widths are found to be $2.894\pm0.011\gev$ and $0.121\pm0.020\gev$ for the $\Zzz$ state, and $2.922\pm0.012\gev$ and $0.138\pm0.029\gev$ for the $\Zzpp$ state. 
The fit parameters are consistent with the earlier result, as expected given the isospin symmetry of the decays.

To estimate the significance of the new $\Zz$ state, pseudoexperiments are generated without the state, and fitted both with and without the $\Zz$ state. The sample size of each pseudoexperiment is Poisson-fluctuated around the number of the observed candidates. Events are generated in the six channels separately, modeled by their individual background and efficiency maps. 
The doubled difference of the log-likelihood $2{\Delta \rm LL}$ of the two fit results should follow a \chisq distribution, where the number of degrees of freedom $N_{\rm df}$ is a fit parameter.
Using 500 pseudoexperiments, $N_{\rm df}$ is determined to be $6.99\pm0.17$. In the obtained \chisq distribution, the value of $2{\Delta \rm LL}$ from collision data corresponds to a significance greater than 9 standard deviations ($\sigma$). 

Among other tested $J^P$ hypotheses beyond the default $0^+$, only the $1^-$ $\Ds\pi$ state leads to a large significance of $6.3\,\sigma$ ($N_{\rm df}=6.99$), while the other cases are not significant. A test of the hypothesis with the simultaneous presence of a $0^+$ $\Zz$ state and a $1^-$ $\Ds\pi$ state yields a significance of $1.3\,\sigma$ ($N_{\rm df}=6.99$), suggesting there is no need in this analysis to include it. Additional $D\pi$, $\Ds\pi$ and $D\Ds$ resonances with spin-parity hypotheses up to $4^+$ are tested with and without the $\Zz$ exotic states. 
In all cases the additional contributions are disfavored, while the $\Zz$ significance remains above $5\,\sigma$ with mass and width parameters consistent with the default results.

\begin{figure}[!tb]
  \begin{center}
    \includegraphics[width=0.5\linewidth]{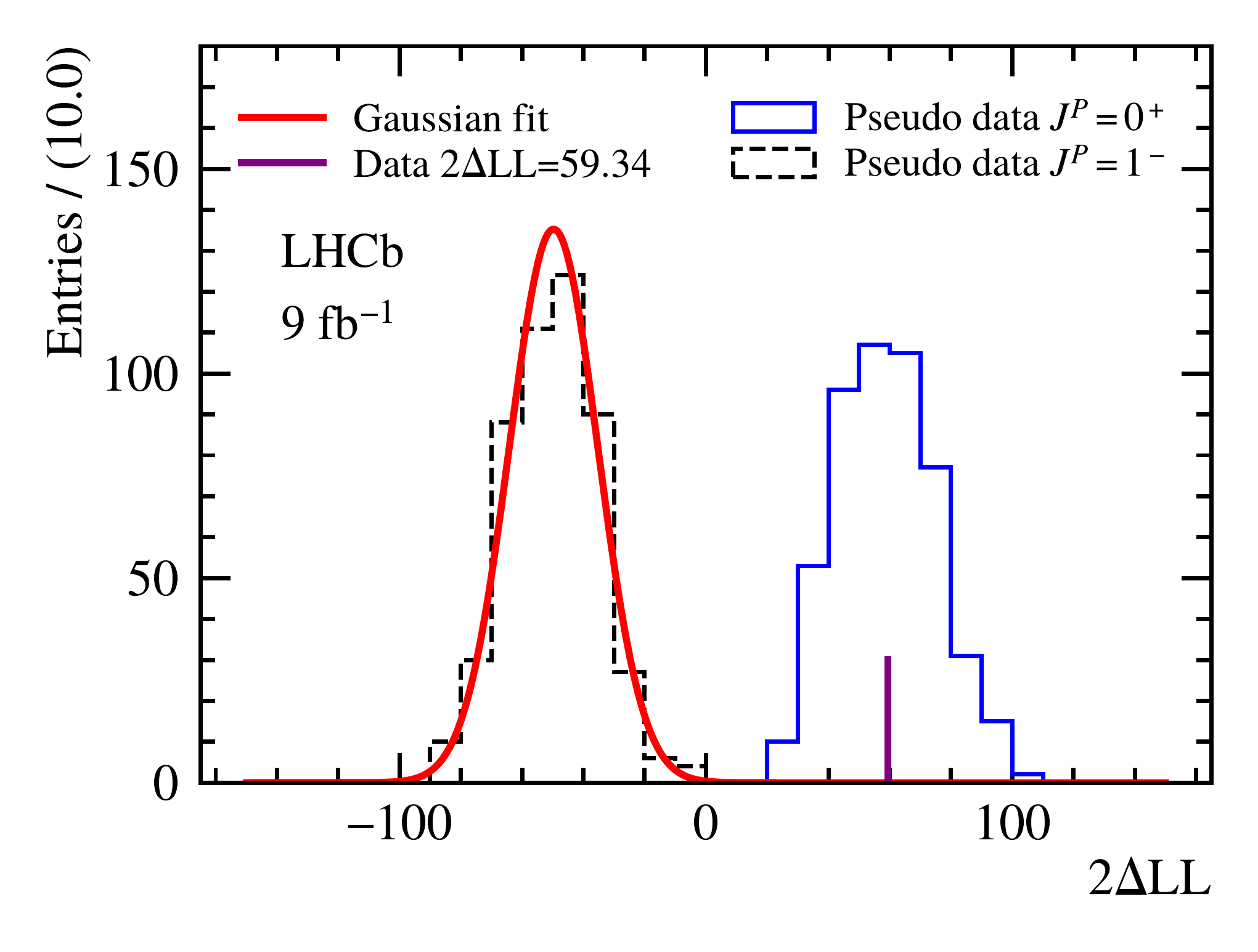}
  \end{center}
  \caption{Spin analysis for the $J^P=0^+$ hypothesis over $J^P=1^-$. The solid blue and black dashed histograms are the distributions of the $2\Delta{\rm LL}$ of the pseudoexperiments based on the fit results with the $0^+$ and $1^+$ hypotheses, respectively. The purple vertical line shows the $2\Delta{\rm LL}$ of the data fitted with the new $\Ds\pi$ exotic state under the $J^P=0^+$ and $J^P=1^-$ hypotheses. The red curve shows the result of a fit with the Gaussian function to the black dashed histogram.}
  \label{fig:spinana}
\end{figure}

Pseudoexperiments are performed to confirm the spin-parity of the $\Zz$ state. A set of 500 pseudoexperiments is generated from the default fit results, another 500 pseudoexperiments are generated from the fit results with a $1^-$ $\Ds\pi$ exotic state included.
Each pseudoexperiment distribution is fitted both under the $J^P=0^+$ hypothesis and $J^P=1^-$ hypothesis, and the $2\Delta{\rm LL}$ are evaluated, shown in Fig.~\ref{fig:spinana}. The $2\Delta{\rm LL}$ value in data corresponds to a significance of about $7.5\,\sigma$. This demonstrates that the $J^P=0^+$ hypothesis for the $\Zz$ state is favored with a high significance.

\begin{figure}[!tb]
  \begin{center}
    \includegraphics[width=0.5\linewidth]{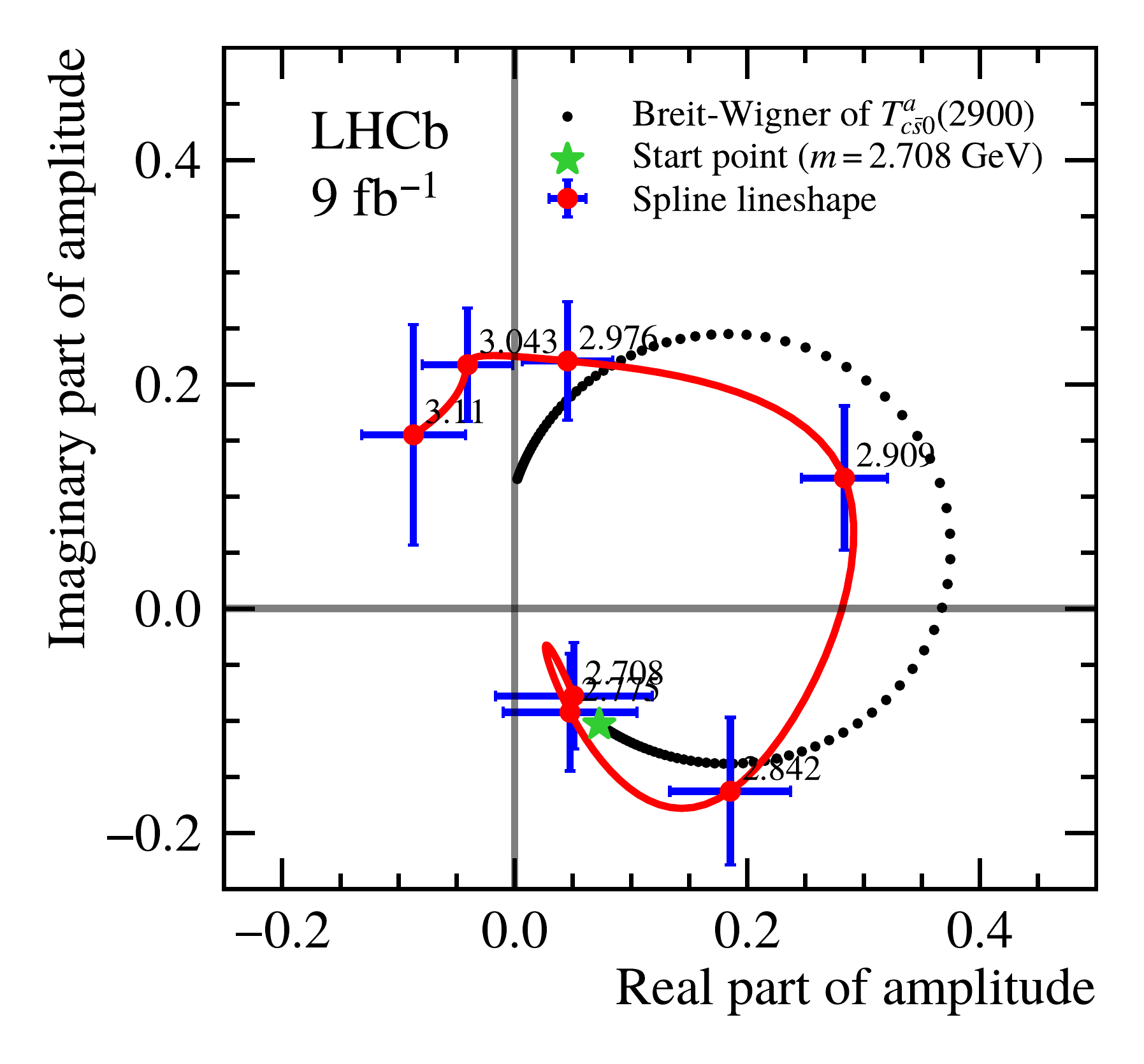}
  \end{center}
  \caption{Argand diagram for the $\Zz$ exotic state. The green star and black dots show the start point and phase variation for a $\Zz$ Breit--Wigner function. The red solid line shows the fitted spline function used to model the $J^P = 0^+$ $\Ds\pi$ component in the $\Zz$ mass region, and the blue error bars show the uncertainties on the values of the spline at the knots.}
  \label{fig:Z0_Argand}
\end{figure}

The Argand diagram~\cite{PDG2022} of the $\Zz$ state, which is expected to have a circular form, is shown in Fig.~\ref{fig:Z0_Argand}. An additional fit, in which the Breit--Wigner lineshape of the $\Zz$ state is replaced by seven spline points on $M(\Ds\pi)$, is performed. The complex factor associated to each point is free to vary. 
The lineshape of the $\Ds\pi$ spline model, also shown in Fig.~\ref{fig:Z0_Argand}, is consistent with the Breit--Wigner lineshape, and further supports the resonant character of the $\Zz$ state.

The sources of systematic uncertainty in the amplitude analysis fall into four categories: the $B$ meson signal yields, the background model, the efficiency map, and the fixed parameters in the fit. 
The dominant systematic uncertainties arise from the fixed Blatt-Weisskopf radius and $D^*$ resonant parameters. Systematic uncertainties due to the Blatt-Weisskopf radius are estimated by varying the parameter from $3.0\gev^{-1}$ to $1.5\gev^{-1}$ and $4.5\gev^{-1}$. Systematic uncertainties associated with the fixed $D^*$ parameters include those related to the masses and widths of the $D^*$ resonances, which are allowed to vary but are constrained to their measured values according to their uncertainties. 
The spline points are shifted by $\pm10\mev$, and the number of spline points is varied by $\pm1$ to explore their impact.
The $J^P$ hypothesis of the $\Dbar(3000)$ state is changed to $2^+$, the next most probable $J^P$ hypothesis, in the systematic uncertainty study. 
The total systematic uncertainty is determined by combining all contributions in quadrature. 
The minimal significance during the variation of the sources of systematic uncertainties is taken as the final significance of the resonance.
Possible biases of fit parameters are studied with pseudoexperiments. A set of 500 pseudoexperiments is generated with the default fit results, and fitted with the same model. The distribution of each fit parameter is fitted with a Gaussian function. Both the residual $(\mu_{\rm pseudo}-\mu_{\rm default})$ and pull distributions $(\mu_{\rm pseudo}-\mu_{\rm default})/\sigma_{\rm pseudo}$ are obtained from the pseudoexperiments, and used to correct the mean values and uncertainties of the fit results.

In summary, a combined amplitude analysis of the \BztoDzbarDsppim and \mbox{\BptoDmDsppip} decays is performed, under the assumption of isospin symmetry, to determine the contributions of intermediate states in the decays. 
Considering only resonant contributions of $D^*$ mesons, the $M(D\pi)$ distributions in the two $B$ decay channels are well modelled, but in the $M(D_s\pi)$ projection of each channel, peaking structures near $2.9\gev$ are not well described.
Two new $D_s\pi$ exotic resonances are included in the fit model. Assuming the neutral $D_s^+\pi^-$ resonance and the doubly charged $D_s^+\pi^+$ resonance belong to the same isospin triplet, 
the common mass and width
are determined to be
\begin{equation}
    \begin{split}
        M=&~2.908\pm0.011\pm0.020\gev,\\
        \Gamma=&~0.136\pm0.023\pm0.013\gev,\nonumber
    \end{split}
\end{equation}
where the first and second uncertainties are statistical and systematic, respectively. The spin-parity is determined to be $0^+$ with a significance of about $7.5\,\sigma$ with respect to the $1^-$ hypothesis. After considering the systematic uncertainties, the significance of the $\Zz$ state is estimated to be greater than $9\,\sigma$, taking into account the look-elsewhere effect.
Meanwhile, amplitude fits with separate $\Zzz$ and $\Zzpp$ resonance parameters are also performed. The masses and widths of the two resonances are
\begin{equation}
    \begin{split}
    \Zzz:   M=&~2.892\pm0.014\pm0.015\gev,\\
            \Gamma=&~0.119\pm0.026\pm0.013\gev,\\
    \Zzpp:  M=&~2.921\pm0.017\pm0.020\gev,\\
            \Gamma=&~0.137\pm0.032\pm0.017\gev,\nonumber
    \end{split}
\end{equation}
which are in good agreement. The significance of the two resonances is found to be $8.0\,\sigma$ $(N_{\rm df}=7.29)$ for the $\Zzz$ state and $6.5\,\sigma$ $(N_{\rm df}=8.53)$ for the $\Zzpp$ state, including systematic effects.
The additional details are described in Ref.~\cite{PRD_twin}.
This is the first observation of a doubly charged tetraquark candidate, $\Zzpp([c\bar{s}u\bar{d}])$, and of its neutral isospin partner, $\Zzz([c\bar{s}\bar{u}d])$. They belong to a new type of open-charm tetraquark states with $c$ and $\bar{s}$ quarks. The obtained mass of the $\Zz$ state is consistent with that of another $0^+$ open-charm tetraquark, the $X_0(2900)^0([cs\bar{u}\bar{d}])$ state discovered in the $D^+K^-$ final state~\cite{LHCb-PAPER-2020-024,LHCb-PAPER-2020-025}, but their widths and flavor contents are different. 
The observation of the $\Zz$ states is a significant step in the study of exotic hadron spectroscopy.

We express our gratitude to our colleagues in the CERN
accelerator departments for the excellent performance of the LHC. We
thank the technical and administrative staff at the LHCb
institutes.
We acknowledge support from CERN and from the national agencies:
CAPES, CNPq, FAPERJ and FINEP (Brazil); 
MOST and NSFC (China); 
CNRS/IN2P3 (France); 
BMBF, DFG and MPG (Germany); 
INFN (Italy); 
NWO (Netherlands); 
MNiSW and NCN (Poland); 
MEN/IFA (Romania); 
MICINN (Spain); 
SNSF and SER (Switzerland); 
NASU (Ukraine); 
STFC (United Kingdom); 
DOE NP and NSF (USA).
We acknowledge the computing resources that are provided by CERN, IN2P3
(France), KIT and DESY (Germany), INFN (Italy), SURF (Netherlands),
PIC (Spain), GridPP (United Kingdom), 
CSCS (Switzerland), IFIN-HH (Romania), CBPF (Brazil),
Polish WLCG  (Poland) and NERSC (USA).
We are indebted to the communities behind the multiple open-source
software packages on which we depend.
Individual groups or members have received support from
ARC and ARDC (Australia);
Minciencias (Colombia);
AvH Foundation (Germany);
EPLANET, Marie Sk\l{}odowska-Curie Actions and ERC (European Union);
A*MIDEX, ANR, IPhU and Labex P2IO, and R\'{e}gion Auvergne-Rh\^{o}ne-Alpes (France);
Key Research Program of Frontier Sciences of CAS, CAS PIFI, CAS CCEPP, 
Fundamental Research Funds for the Central Universities, 
and Sci. \& Tech. Program of Guangzhou (China);
GVA, XuntaGal, GENCAT and Prog.~Atracci\'on Talento, CM (Spain);
SRC (Sweden);
the Leverhulme Trust, the Royal Society
 and UKRI (United Kingdom).

\addcontentsline{toc}{section}{References}
\ifx\mcitethebibliography\mciteundefinedmacro
\PackageError{LHCb.bst}{mciteplus.sty has not been loaded}
{This bibstyle requires the use of the mciteplus package.}\fi
\providecommand{\href}[2]{#2}

\newpage
\centerline
{\large\bf LHCb collaboration}
\begin
{flushleft}
\small
R.~Aaij$^{32}$\lhcborcid{0000-0003-0533-1952},
A.S.W.~Abdelmotteleb$^{50}$\lhcborcid{0000-0001-7905-0542},
C.~Abellan~Beteta$^{44}$,
F.~Abudin{\'e}n$^{50}$\lhcborcid{0000-0002-6737-3528},
T.~Ackernley$^{54}$\lhcborcid{0000-0002-5951-3498},
B.~Adeva$^{40}$\lhcborcid{0000-0001-9756-3712},
M.~Adinolfi$^{48}$\lhcborcid{0000-0002-1326-1264},
P.~Adlarson$^{77}$\lhcborcid{0000-0001-6280-3851},
H.~Afsharnia$^{9}$,
C.~Agapopoulou$^{13}$\lhcborcid{0000-0002-2368-0147}
% ...

{\footnotesize \it

% ...
$^{9}$Universit{\'e} Clermont Auvergne, CNRS/IN2P3, LPC, Clermont-Ferrand, France\\
% ...
$^{13}$LPNHE, Sorbonne Universit{\'e}, Paris Diderot Sorbonne Paris Cit{\'e}, CNRS/IN2P3, Paris, France\\
% ...
$^{32}$Nikhef National Institute for Subatomic Physics, Amsterdam, Netherlands\\
% ...
$^{40}$Instituto Galego de F{\'\i}sica de Altas Enerx{\'\i}as (IGFAE), Universidade de Santiago de Compostela, Santiago de Compostela, Spain\\
% ...
$^{44}$Physik-Institut, Universit{\"a}t Z{\"u}rich, Z{\"u}rich, Switzerland\\
% ...
$^{48}$H.H. Wills Physics Laboratory, University of Bristol, Bristol, United Kingdom\\
% ...
$^{50}$Department of Physics, University of Warwick, Coventry, United Kingdom\\
% ...
$^{53}$School of Physics and Astronomy, University of Glasgow, Glasgow, United Kingdom\\
$^{54}$Oliver Lodge Laboratory, University of Liverpool, Liverpool, United Kingdom\\
% ...
$^{77}$Department of Physics and Astronomy, Uppsala University, Uppsala, Sweden, associated to $^{53}$\\
% ...
}
\end{flushleft}


\begin{mcitethebibliography}{10}
\mciteSetBstSublistMode{n}
\mciteSetBstMaxWidthForm{subitem}{\alph{mcitesubitemcount})}
\mciteSetBstSublistLabelBeginEnd{\mcitemaxwidthsubitemform\space}
{\relax}{\relax}

\bibitem{Gell-Mann:1964ewy}
M.~Gell-Mann, \ifthenelse{\boolean{articletitles}}{\emph{{A schematic model of
  baryons and mesons}},
  }{}\href{https://doi.org/10.1016/S0031-9163(64)92001-3}{Phys.\ Lett.\
  \textbf{8} (1964) 214}\relax
\mciteBstWouldAddEndPuncttrue
\mciteSetBstMidEndSepPunct{\mcitedefaultmidpunct}
{\mcitedefaultendpunct}{\mcitedefaultseppunct}\relax
\EndOfBibitem
\bibitem{Molina:2010tx}
R.~Molina, T.~Branz, and E.~Oset, \ifthenelse{\boolean{articletitles}}{\emph{{A
  new interpretation for the $D^*_{s2}(2573)$ and the prediction of novel
  exotic charmed mesons}},
  }{}\href{https://doi.org/10.1103/PhysRevD.82.014010}{Phys.\ Rev.\
  \textbf{D82} (2010) 014010},
  \href{http://arxiv.org/abs/1005.0335}{{\normalfont\ttfamily
  arXiv:1005.0335}}\relax
\mciteBstWouldAddEndPuncttrue
\mciteSetBstMidEndSepPunct{\mcitedefaultmidpunct}
{\mcitedefaultendpunct}{\mcitedefaultseppunct}\relax
\EndOfBibitem
\bibitem{Zhu:2022arXiv}
H.-X. Chen {\em et~al.}, \ifthenelse{\boolean{articletitles}}{\emph{{An updated
  review of the new hadron states}},
  }{}\href{http://arxiv.org/abs/2204.02649}{{\normalfont\ttfamily
  arXiv:2204.02649}}\relax
\mciteBstWouldAddEndPuncttrue
\mciteSetBstMidEndSepPunct{\mcitedefaultmidpunct}
{\mcitedefaultendpunct}{\mcitedefaultseppunct}\relax
\EndOfBibitem
\bibitem{BaBar:2003oey}
BaBar collaboration, B.~Aubert {\em et~al.},
  \ifthenelse{\boolean{articletitles}}{\emph{{Observation of a narrow meson
  decaying to $D_s^+ \pi^0$ at a mass of 2.32-GeV/c$^2$}},
  }{}\href{https://doi.org/10.1103/PhysRevLett.90.242001}{Phys.\ Rev.\ Lett.\
  \textbf{90} (2003) 242001},
  \href{http://arxiv.org/abs/hep-ex/0304021}{{\normalfont\ttfamily
  arXiv:hep-ex/0304021}}\relax
\mciteBstWouldAddEndPuncttrue
\mciteSetBstMidEndSepPunct{\mcitedefaultmidpunct}
{\mcitedefaultendpunct}{\mcitedefaultseppunct}\relax
\EndOfBibitem
\bibitem{CLEO:2003ggt}
CLEO collaboration, D.~Besson {\em et~al.},
  \ifthenelse{\boolean{articletitles}}{\emph{{Observation of a narrow resonance
  of mass 2.46-GeV/c**2 decaying to D*+(s) pi0 and confirmation of the
  D*(sJ)(2317) state}},
  }{}\href{https://doi.org/10.1103/PhysRevD.68.032002}{Phys.\ Rev.\ D
  \textbf{68} (2003) 032002},
  \href{http://arxiv.org/abs/hep-ex/0305100}{{\normalfont\ttfamily
  arXiv:hep-ex/0305100}}, [Erratum: Phys.Rev.D 75, 119908 (2007)]\relax
\mciteBstWouldAddEndPuncttrue
\mciteSetBstMidEndSepPunct{\mcitedefaultmidpunct}
{\mcitedefaultendpunct}{\mcitedefaultseppunct}\relax
\EndOfBibitem
\bibitem{PDG2022}
Particle Data Group, R.~L. Workman {\em et~al.},
  \ifthenelse{\boolean{articletitles}}{\emph{{\href{http://pdg.lbl.gov/}{Review
  of particle physics}}}, }{}\href{https://doi.org/10.1093/ptep/ptac097}{Prog.\
  Theor.\ Exp.\ Phys.\  \textbf{2022} (2022) 083C01}\relax
\mciteBstWouldAddEndPuncttrue
\mciteSetBstMidEndSepPunct{\mcitedefaultmidpunct}
{\mcitedefaultendpunct}{\mcitedefaultseppunct}\relax
\EndOfBibitem
\bibitem{Chen:2016spr}
H.-X. Chen {\em et~al.}, \ifthenelse{\boolean{articletitles}}{\emph{{A review
  of the open charm and open bottom systems}},
  }{}\href{https://doi.org/10.1088/1361-6633/aa6420}{Rept.\ Prog.\ Phys.\
  \textbf{80} (2017) 076201},
  \href{http://arxiv.org/abs/1609.08928}{{\normalfont\ttfamily
  arXiv:1609.08928}}\relax
\mciteBstWouldAddEndPuncttrue
\mciteSetBstMidEndSepPunct{\mcitedefaultmidpunct}
{\mcitedefaultendpunct}{\mcitedefaultseppunct}\relax
\EndOfBibitem
\bibitem{BaBar:2006eep}
BaBar collaboration, B.~Aubert {\em et~al.},
  \ifthenelse{\boolean{articletitles}}{\emph{{A Study of the D*(sJ)(2317) and
  D(sJ)(2460) Mesons in Inclusive c anti-c Production Near (s)**(1/2) =
  10.6-GeV}}, }{}\href{https://doi.org/10.1103/PhysRevD.74.032007}{Phys.\ Rev.\
  D \textbf{74} (2006) 032007},
  \href{http://arxiv.org/abs/hep-ex/0604030}{{\normalfont\ttfamily
  arXiv:hep-ex/0604030}}\relax
\mciteBstWouldAddEndPuncttrue
\mciteSetBstMidEndSepPunct{\mcitedefaultmidpunct}
{\mcitedefaultendpunct}{\mcitedefaultseppunct}\relax
\EndOfBibitem
\bibitem{Belle:2015glz}
Belle collaboration, S.-K. Choi {\em et~al.},
  \ifthenelse{\boolean{articletitles}}{\emph{{Measurements of $B\rightarrow
  \bar{D} D_{s0} ^{*+}(2317)$ decay rates and a search for isospin partners of
  the $D_{s0}^{*+} (2317)$}},
  }{}\href{https://doi.org/10.1103/PhysRevD.91.092011}{Phys.\ Rev.\ D
  \textbf{91} (2015) 092011},
  \href{http://arxiv.org/abs/1504.02637}{{\normalfont\ttfamily
  arXiv:1504.02637}}, [Addendum: Phys.Rev.D 92, 039905 (2015)]\relax
\mciteBstWouldAddEndPuncttrue
\mciteSetBstMidEndSepPunct{\mcitedefaultmidpunct}
{\mcitedefaultendpunct}{\mcitedefaultseppunct}\relax
\EndOfBibitem
\bibitem{D0:2016mwd}
D0 collaboration, V.~M. Abazov {\em et~al.},
  \ifthenelse{\boolean{articletitles}}{\emph{{Evidence for a $B_s^0 \pi^\pm$
  state}}, }{}\href{https://doi.org/10.1103/PhysRevLett.117.022003}{Phys.\
  Rev.\ Lett.\  \textbf{117} (2016) 022003},
  \href{http://arxiv.org/abs/1602.07588}{{\normalfont\ttfamily
  arXiv:1602.07588}}\relax
\mciteBstWouldAddEndPuncttrue
\mciteSetBstMidEndSepPunct{\mcitedefaultmidpunct}
{\mcitedefaultendpunct}{\mcitedefaultseppunct}\relax
\EndOfBibitem
\bibitem{D0:2017qqm}
D0 collaboration, V.~M. Abazov {\em et~al.},
  \ifthenelse{\boolean{articletitles}}{\emph{{Study of the ${X^\pm(5568)}$
  state with semileptonic decays of the ${B_s^0}$ meson}},
  }{}\href{https://doi.org/10.1103/PhysRevD.97.092004}{Phys.\ Rev.\
  \textbf{D97} (2018) 092004},
  \href{http://arxiv.org/abs/1712.10176}{{\normalfont\ttfamily
  arXiv:1712.10176}}\relax
\mciteBstWouldAddEndPuncttrue
\mciteSetBstMidEndSepPunct{\mcitedefaultmidpunct}
{\mcitedefaultendpunct}{\mcitedefaultseppunct}\relax
\EndOfBibitem
\bibitem{LHCb-PAPER-2016-029}
LHCb collaboration, R.~Aaij {\em et~al.},
  \ifthenelse{\boolean{articletitles}}{\emph{{Search for structure in the
  $\Bs\pipm$ invariant mass spectrum}},
  }{}\href{https://doi.org/10.1103/PhysRevLett.117.152003}{Phys.\ Rev.\ Lett.\
  \textbf{117} (2016) 152003},
  \href{http://arxiv.org/abs/1608.00435}{{\normalfont\ttfamily
  arXiv:1608.00435}}\relax
\mciteBstWouldAddEndPuncttrue
\mciteSetBstMidEndSepPunct{\mcitedefaultmidpunct}
{\mcitedefaultendpunct}{\mcitedefaultseppunct}\relax
\EndOfBibitem
\bibitem{CMS:2017hfy}
CMS collaboration, A.~M. Sirunyan {\em et~al.},
  \ifthenelse{\boolean{articletitles}}{\emph{{Search for the $X(5568)$ state
  decaying into $\Bs\pipm$ in proton-proton collisions at $\sqrt{s} = 8
  \tev$}}, }{}\href{https://doi.org/10.1103/PhysRevLett.120.202005}{Phys.\
  Rev.\ Lett.\  \textbf{120} (2018) 202005},
  \href{http://arxiv.org/abs/1712.06144}{{\normalfont\ttfamily
  arXiv:1712.06144}}\relax
\mciteBstWouldAddEndPuncttrue
\mciteSetBstMidEndSepPunct{\mcitedefaultmidpunct}
{\mcitedefaultendpunct}{\mcitedefaultseppunct}\relax
\EndOfBibitem
\bibitem{ATLAS:2018udc}
ATLAS collaboration, M.~Aaboud {\em et~al.},
  \ifthenelse{\boolean{articletitles}}{\emph{{Search for a structure in the
  $B^0_s \pi^\pm$ invariant mass spectrum with the ATLAS experiment}},
  }{}\href{https://doi.org/10.1103/PhysRevLett.120.202007}{Phys.\ Rev.\ Lett.\
  \textbf{120} (2018) 202007},
  \href{http://arxiv.org/abs/1802.01840}{{\normalfont\ttfamily
  arXiv:1802.01840}}\relax
\mciteBstWouldAddEndPuncttrue
\mciteSetBstMidEndSepPunct{\mcitedefaultmidpunct}
{\mcitedefaultendpunct}{\mcitedefaultseppunct}\relax
\EndOfBibitem
\bibitem{CDF:2017dwr}
CDF collaboration, T.~Aaltonen {\em et~al.},
  \ifthenelse{\boolean{articletitles}}{\emph{{A search for the exotic meson
  $X(5568)$ with the Collider Detector at Fermilab}},
  }{}\href{https://doi.org/10.1103/PhysRevLett.120.202006}{Phys.\ Rev.\ Lett.\
  \textbf{120} (2018) 202006},
  \href{http://arxiv.org/abs/1712.09620}{{\normalfont\ttfamily
  arXiv:1712.09620}}\relax
\mciteBstWouldAddEndPuncttrue
\mciteSetBstMidEndSepPunct{\mcitedefaultmidpunct}
{\mcitedefaultendpunct}{\mcitedefaultseppunct}\relax
\EndOfBibitem
\bibitem{Agaev:2016lkl}
S.~S. Agaev, K.~Azizi, and H.~Sundu,
  \ifthenelse{\boolean{articletitles}}{\emph{{Charmed partner of the exotic
  $X(5568)$ state and its properties}},
  }{}\href{https://doi.org/10.1103/PhysRevD.93.094006}{Phys.\ Rev.\
  \textbf{D93} (2016) 094006},
  \href{http://arxiv.org/abs/1603.01471}{{\normalfont\ttfamily
  arXiv:1603.01471}}\relax
\mciteBstWouldAddEndPuncttrue
\mciteSetBstMidEndSepPunct{\mcitedefaultmidpunct}
{\mcitedefaultendpunct}{\mcitedefaultseppunct}\relax
\EndOfBibitem
\bibitem{Chen:2017rhl}
W.~Chen {\em et~al.}, \ifthenelse{\boolean{articletitles}}{\emph{{Open-flavor
  charm and bottom $sq\bar q\bar Q$ and $qq\bar q\bar Q$ tetraquark states}},
  }{}\href{https://doi.org/10.1103/PhysRevD.95.114005}{Phys.\ Rev.\
  \textbf{D95} (2017) 114005},
  \href{http://arxiv.org/abs/1705.10088}{{\normalfont\ttfamily
  arXiv:1705.10088}}\relax
\mciteBstWouldAddEndPuncttrue
\mciteSetBstMidEndSepPunct{\mcitedefaultmidpunct}
{\mcitedefaultendpunct}{\mcitedefaultseppunct}\relax
\EndOfBibitem
\bibitem{Baeza-Ballesteros:2022azb}
J.~Baeza-Ballesteros, P.~Hern\'andez, and F.~Romero-L\'opez,
  \ifthenelse{\boolean{articletitles}}{\emph{{A lattice study of
  \ensuremath{\pi}\ensuremath{\pi} scattering at large N$_{c}$}},
  }{}\href{https://doi.org/10.1007/JHEP06(2022)049}{JHEP \textbf{06} (2022)
  049}, \href{http://arxiv.org/abs/2202.02291}{{\normalfont\ttfamily
  arXiv:2202.02291}}\relax
\mciteBstWouldAddEndPuncttrue
\mciteSetBstMidEndSepPunct{\mcitedefaultmidpunct}
{\mcitedefaultendpunct}{\mcitedefaultseppunct}\relax
\EndOfBibitem
\bibitem{LHCb-PAPER-2020-024}
LHCb collaboration, R.~Aaij {\em et~al.},
  \ifthenelse{\boolean{articletitles}}{\emph{{Model-independent study of
  structure in \mbox{$\Bp \to \Dp \Dm \Kp$} decays}},
  }{}\href{https://doi.org/10.1103/PhysRevLett.125.242001}{Phys.\ Rev.\ Lett.\
  \textbf{125} (2020) 242001},
  \href{http://arxiv.org/abs/2009.00025}{{\normalfont\ttfamily
  arXiv:2009.00025}}\relax
\mciteBstWouldAddEndPuncttrue
\mciteSetBstMidEndSepPunct{\mcitedefaultmidpunct}
{\mcitedefaultendpunct}{\mcitedefaultseppunct}\relax
\EndOfBibitem
\bibitem{LHCb-PAPER-2020-025}
LHCb collaboration, R.~Aaij {\em et~al.},
  \ifthenelse{\boolean{articletitles}}{\emph{{Amplitude analysis of the $\Bp
  \to \Dp \Dm \Kp$ decay}},
  }{}\href{https://doi.org/10.1103/PhysRevD.102.112003}{Phys.\ Rev.\
  \textbf{D102} (2020) 112003},
  \href{http://arxiv.org/abs/2009.00026}{{\normalfont\ttfamily
  arXiv:2009.00026}}\relax
\mciteBstWouldAddEndPuncttrue
\mciteSetBstMidEndSepPunct{\mcitedefaultmidpunct}
{\mcitedefaultendpunct}{\mcitedefaultseppunct}\relax
\EndOfBibitem
\bibitem{He:2020jna}
X.-G. He, W.~Wang, and R.~Zhu,
  \ifthenelse{\boolean{articletitles}}{\emph{{Open-charm tetraquark $X_c$ and
  open-bottom tetraquark $X_b$}},
  }{}\href{https://doi.org/10.1140/epjc/s10052-020-08597-1}{Eur.\ Phys.\ J.\
  \textbf{C80} (2020) 1026},
  \href{http://arxiv.org/abs/2008.07145}{{\normalfont\ttfamily
  arXiv:2008.07145}}\relax
\mciteBstWouldAddEndPuncttrue
\mciteSetBstMidEndSepPunct{\mcitedefaultmidpunct}
{\mcitedefaultendpunct}{\mcitedefaultseppunct}\relax
\EndOfBibitem
\bibitem{Lu:2020qmp}
Q.-F. L\"u, D.-Y. Chen, and Y.-B. Dong,
  \ifthenelse{\boolean{articletitles}}{\emph{{Open charm and bottom tetraquarks
  in an extended relativized quark model}},
  }{}\href{https://doi.org/10.1103/PhysRevD.102.074021}{Phys.\ Rev.\
  \textbf{C102} (2020) 074021},
  \href{http://arxiv.org/abs/2008.07340}{{\normalfont\ttfamily
  arXiv:2008.07340}}\relax
\mciteBstWouldAddEndPuncttrue
\mciteSetBstMidEndSepPunct{\mcitedefaultmidpunct}
{\mcitedefaultendpunct}{\mcitedefaultseppunct}\relax
\EndOfBibitem
\bibitem{Burns:2020xne}
T.~J. Burns and E.~S. Swanson,
  \ifthenelse{\boolean{articletitles}}{\emph{{Discriminating among
  interpretations for the $X(2900)$ states}},
  }{}\href{https://doi.org/10.1103/PhysRevD.103.014004}{Phys.\ Rev.\
  \textbf{D103} (2021) 014004},
  \href{http://arxiv.org/abs/2009.05352}{{\normalfont\ttfamily
  arXiv:2009.05352}}\relax
\mciteBstWouldAddEndPuncttrue
\mciteSetBstMidEndSepPunct{\mcitedefaultmidpunct}
{\mcitedefaultendpunct}{\mcitedefaultseppunct}\relax
\EndOfBibitem
\bibitem{Agaev:2021knl}
S.~S. Agaev, K.~Azizi, and H.~Sundu,
  \ifthenelse{\boolean{articletitles}}{\emph{{Vector resonance $X_1$(2900) and
  its structure}},
  }{}\href{https://doi.org/10.1016/j.nuclphysa.2021.122202}{Nucl.\ Phys.\
  \textbf{A1011} (2021) 122202},
  \href{http://arxiv.org/abs/2103.06151}{{\normalfont\ttfamily
  arXiv:2103.06151}}\relax
\mciteBstWouldAddEndPuncttrue
\mciteSetBstMidEndSepPunct{\mcitedefaultmidpunct}
{\mcitedefaultendpunct}{\mcitedefaultseppunct}\relax
\EndOfBibitem
\bibitem{Agaev:2021jsz}
S.~S. Agaev, K.~Azizi, and H.~Sundu,
  \ifthenelse{\boolean{articletitles}}{\emph{{Doubly charged vector tetraquark
  $Z_V^{++}=[cu][\bar{s}\bar{d}]$}},
  }{}\href{https://doi.org/10.1016/j.physletb.2021.136530}{Phys.\ Lett.\
  \textbf{B820} (2021) 136530},
  \href{http://arxiv.org/abs/2105.00081}{{\normalfont\ttfamily
  arXiv:2105.00081}}\relax
\mciteBstWouldAddEndPuncttrue
\mciteSetBstMidEndSepPunct{\mcitedefaultmidpunct}
{\mcitedefaultendpunct}{\mcitedefaultseppunct}\relax
\EndOfBibitem
\bibitem{Azizi:2021aib}
K.~Azizi and U.~\"Ozdem, \ifthenelse{\boolean{articletitles}}{\emph{{Magnetic
  dipole moments of the $T_{cc}^+$ and $Z_V^{++}$ tetraquark states}},
  }{}\href{https://doi.org/10.1103/PhysRevD.104.114002}{Phys.\ Rev.\
  \textbf{D104} (2021) 114002},
  \href{http://arxiv.org/abs/2109.02390}{{\normalfont\ttfamily
  arXiv:2109.02390}}\relax
\mciteBstWouldAddEndPuncttrue
\mciteSetBstMidEndSepPunct{\mcitedefaultmidpunct}
{\mcitedefaultendpunct}{\mcitedefaultseppunct}\relax
\EndOfBibitem
\bibitem{LHCB-PAPER-2013-026}
LHCb collaboration, R.~Aaij {\em et~al.},
  \ifthenelse{\boolean{articletitles}}{\emph{{Study of $D_J$ meson decays to
  $\Dp\pim$, $\Dz\pip$ and $\Dstarp\pim$ final states in \proton\proton
  collisions}}, }{}\href{https://doi.org/10.1007/JHEP09(2013)145}{JHEP
  \textbf{09} (2013) 145},
  \href{http://arxiv.org/abs/1307.4556}{{\normalfont\ttfamily
  arXiv:1307.4556}}\relax
\mciteBstWouldAddEndPuncttrue
\mciteSetBstMidEndSepPunct{\mcitedefaultmidpunct}
{\mcitedefaultendpunct}{\mcitedefaultseppunct}\relax
\EndOfBibitem
\bibitem{LHCB-PAPER-2014-070}
LHCb collaboration, R.~Aaij {\em et~al.},
  \ifthenelse{\boolean{articletitles}}{\emph{{Dalitz plot analysis of
  \mbox{\decay{\Bz}{\Dzb\pip\pim}} decays}},
  }{}\href{https://doi.org/10.1103/PhysRevD.92.032002}{Phys.\ Rev.\
  \textbf{D92} (2015) 032002},
  \href{http://arxiv.org/abs/1505.01710}{{\normalfont\ttfamily
  arXiv:1505.01710}}\relax
\mciteBstWouldAddEndPuncttrue
\mciteSetBstMidEndSepPunct{\mcitedefaultmidpunct}
{\mcitedefaultendpunct}{\mcitedefaultseppunct}\relax
\EndOfBibitem
\bibitem{LHCB-PAPER-2016-026}
LHCb collaboration, R.~Aaij {\em et~al.},
  \ifthenelse{\boolean{articletitles}}{\emph{{Amplitude analysis of
  \mbox{\decay{\Bm}{\Dp \pim \pim}} decays}},
  }{}\href{https://doi.org/10.1103/PhysRevD.94.072001}{Phys.\ Rev.\
  \textbf{D94} (2016) 072001},
  \href{http://arxiv.org/abs/1608.01289}{{\normalfont\ttfamily
  arXiv:1608.01289}}\relax
\mciteBstWouldAddEndPuncttrue
\mciteSetBstMidEndSepPunct{\mcitedefaultmidpunct}
{\mcitedefaultendpunct}{\mcitedefaultseppunct}\relax
\EndOfBibitem
\bibitem{LHCb-DP-2008-001}
LHCb collaboration, A.~A. Alves~Jr.\ {\em et~al.},
  \ifthenelse{\boolean{articletitles}}{\emph{{The \lhcb detector at the LHC}},
  }{}\href{https://doi.org/10.1088/1748-0221/3/08/S08005}{JINST \textbf{3}
  (2008) S08005}\relax
\mciteBstWouldAddEndPuncttrue
\mciteSetBstMidEndSepPunct{\mcitedefaultmidpunct}
{\mcitedefaultendpunct}{\mcitedefaultseppunct}\relax
\EndOfBibitem
\bibitem{LHCb-DP-2014-002}
LHCb collaboration, R.~Aaij {\em et~al.},
  \ifthenelse{\boolean{articletitles}}{\emph{{LHCb detector performance}},
  }{}\href{https://doi.org/10.1142/S0217751X15300227}{Int.\ J.\ Mod.\ Phys.\
  \textbf{A30} (2015) 1530022},
  \href{http://arxiv.org/abs/1412.6352}{{\normalfont\ttfamily
  arXiv:1412.6352}}\relax
\mciteBstWouldAddEndPuncttrue
\mciteSetBstMidEndSepPunct{\mcitedefaultmidpunct}
{\mcitedefaultendpunct}{\mcitedefaultseppunct}\relax
\EndOfBibitem
\bibitem{Sjostrand:2007gs}
T.~Sj\"{o}strand, S.~Mrenna, and P.~Skands,
  \ifthenelse{\boolean{articletitles}}{\emph{{A brief introduction to PYTHIA
  8.1}}, }{}\href{https://doi.org/10.1016/j.cpc.2008.01.036}{Comput.\ Phys.\
  Commun.\  \textbf{178} (2008) 852},
  \href{http://arxiv.org/abs/0710.3820}{{\normalfont\ttfamily
  arXiv:0710.3820}}\relax
\mciteBstWouldAddEndPuncttrue
\mciteSetBstMidEndSepPunct{\mcitedefaultmidpunct}
{\mcitedefaultendpunct}{\mcitedefaultseppunct}\relax
\EndOfBibitem
\bibitem{Sjostrand:2006za}
T.~Sj\"{o}strand, S.~Mrenna, and P.~Skands,
  \ifthenelse{\boolean{articletitles}}{\emph{{PYTHIA 6.4 physics and manual}},
  }{}\href{https://doi.org/10.1088/1126-6708/2006/05/026}{JHEP \textbf{05}
  (2006) 026}, \href{http://arxiv.org/abs/hep-ph/0603175}{{\normalfont\ttfamily
  arXiv:hep-ph/0603175}}\relax
\mciteBstWouldAddEndPuncttrue
\mciteSetBstMidEndSepPunct{\mcitedefaultmidpunct}
{\mcitedefaultendpunct}{\mcitedefaultseppunct}\relax
\EndOfBibitem
\bibitem{Lange:2001uf}
D.~J. Lange, \ifthenelse{\boolean{articletitles}}{\emph{{The EvtGen particle
  decay simulation package}},
  }{}\href{https://doi.org/10.1016/S0168-9002(01)00089-4}{Nucl.\ Instrum.\
  Meth.\  \textbf{A462} (2001) 152}\relax
\mciteBstWouldAddEndPuncttrue
\mciteSetBstMidEndSepPunct{\mcitedefaultmidpunct}
{\mcitedefaultendpunct}{\mcitedefaultseppunct}\relax
\EndOfBibitem
\bibitem{Allison:2006ve}
Geant4 collaboration, J.~Allison {\em et~al.},
  \ifthenelse{\boolean{articletitles}}{\emph{{Geant4 developments and
  applications}}, }{}\href{https://doi.org/10.1109/TNS.2006.869826}{IEEE
  Trans.\ Nucl.\ Sci.\  \textbf{53} (2006) 270}\relax
\mciteBstWouldAddEndPuncttrue
\mciteSetBstMidEndSepPunct{\mcitedefaultmidpunct}
{\mcitedefaultendpunct}{\mcitedefaultseppunct}\relax
\EndOfBibitem
\bibitem{Agostinelli:2002hh}
Geant4 collaboration, S.~Agostinelli {\em et~al.},
  \ifthenelse{\boolean{articletitles}}{\emph{{Geant4: A simulation toolkit}},
  }{}\href{https://doi.org/10.1016/S0168-9002(03)01368-8}{Nucl.\ Instrum.\
  Meth.\  \textbf{A506} (2003) 250}\relax
\mciteBstWouldAddEndPuncttrue
\mciteSetBstMidEndSepPunct{\mcitedefaultmidpunct}
{\mcitedefaultendpunct}{\mcitedefaultseppunct}\relax
\EndOfBibitem
\bibitem{LHCb-PROC-2011-006}
M.~Clemencic {\em et~al.}, \ifthenelse{\boolean{articletitles}}{\emph{{The
  \lhcb simulation application, Gauss: Design, evolution and experience}},
  }{}\href{https://doi.org/10.1088/1742-6596/331/3/032023}{J.\ Phys.\ Conf.\
  Ser.\  \textbf{331} (2011) 032023}\relax
\mciteBstWouldAddEndPuncttrue
\mciteSetBstMidEndSepPunct{\mcitedefaultmidpunct}
{\mcitedefaultendpunct}{\mcitedefaultseppunct}\relax
\EndOfBibitem
\bibitem{PRD_twin}
LHCb collaboration, R.~Aaij {\em et~al.},
  \ifthenelse{\boolean{articletitles}}{\emph{{Amplitude analysis of $B^0 \to
  \offsetoverline{D}^0 D_s^+ \pi^-$ and $B^+ \to D^- D_s^+ \pi^+$ decays}},
  }{}simultaneous submission to PRD (2022)\relax
\mciteBstWouldAddEndPuncttrue
\mciteSetBstMidEndSepPunct{\mcitedefaultmidpunct}
{\mcitedefaultendpunct}{\mcitedefaultseppunct}\relax
\EndOfBibitem
\bibitem{Breiman}
L.~Breiman, J.~H. Friedman, R.~A. Olshen, and C.~J. Stone, {\em Classification
  and regression trees}, Wadsworth international group, Belmont, California,
  USA, 1984\relax
\mciteBstWouldAddEndPuncttrue
\mciteSetBstMidEndSepPunct{\mcitedefaultmidpunct}
{\mcitedefaultendpunct}{\mcitedefaultseppunct}\relax
\EndOfBibitem
\bibitem{AdaBoost}
Y.~Freund and R.~E. Schapire, \ifthenelse{\boolean{articletitles}}{\emph{A
  decision-theoretic generalization of on-line learning and an application to
  boosting}, }{}\href{https://doi.org/10.1006/jcss.1997.1504}{J.\ Comput.\
  Syst.\ Sci.\  \textbf{55} (1997) 119}\relax
\mciteBstWouldAddEndPuncttrue
\mciteSetBstMidEndSepPunct{\mcitedefaultmidpunct}
{\mcitedefaultendpunct}{\mcitedefaultseppunct}\relax
\EndOfBibitem
\bibitem{Hocker:2007ht}
H.~Voss, A.~Hoecker, J.~Stelzer, and F.~Tegenfeldt,
  \ifthenelse{\boolean{articletitles}}{\emph{{TMVA - Toolkit for Multivariate
  Data Analysis with ROOT}}, }{}\href{https://doi.org/10.22323/1.050.0040}{PoS
  \textbf{ACAT} (2007) 040}\relax
\mciteBstWouldAddEndPuncttrue
\mciteSetBstMidEndSepPunct{\mcitedefaultmidpunct}
{\mcitedefaultendpunct}{\mcitedefaultseppunct}\relax
\EndOfBibitem
\bibitem{TMVA4}
A.~Hoecker {\em et~al.}, \ifthenelse{\boolean{articletitles}}{\emph{{TMVA 4 ---
  Toolkit for Multivariate Data Analysis with ROOT. Users Guide.}},
  }{}\href{http://arxiv.org/abs/physics/0703039}{{\normalfont\ttfamily
  arXiv:physics/0703039}}\relax
\mciteBstWouldAddEndPuncttrue
\mciteSetBstMidEndSepPunct{\mcitedefaultmidpunct}
{\mcitedefaultendpunct}{\mcitedefaultseppunct}\relax
\EndOfBibitem
\bibitem{Poluektov:2014rxa}
A.~Poluektov, \ifthenelse{\boolean{articletitles}}{\emph{{Kernel density
  estimation of a multidimensional efficiency profile}},
  }{}\href{https://doi.org/10.1088/1748-0221/10/02/P02011}{JINST \textbf{10}
  (2015) P02011}, \href{http://arxiv.org/abs/1411.5528}{{\normalfont\ttfamily
  arXiv:1411.5528}}\relax
\mciteBstWouldAddEndPuncttrue
\mciteSetBstMidEndSepPunct{\mcitedefaultmidpunct}
{\mcitedefaultendpunct}{\mcitedefaultseppunct}\relax
\EndOfBibitem
\bibitem{mathad2021efficient}
A.~Mathad, D.~O'Hanlon, A.~Poluektov, and R.~Rabadan,
  \ifthenelse{\boolean{articletitles}}{\emph{{Efficient description of
  experimental effects in amplitude analyses}},
  }{}\href{https://doi.org/10.1088/1748-0221/16/06/P06016}{JINST \textbf{16}
  (2021) P06016}, \href{http://arxiv.org/abs/1902.01452}{{\normalfont\ttfamily
  arXiv:1902.01452}}\relax
\mciteBstWouldAddEndPuncttrue
\mciteSetBstMidEndSepPunct{\mcitedefaultmidpunct}
{\mcitedefaultendpunct}{\mcitedefaultseppunct}\relax
\EndOfBibitem
\bibitem{MySupp}
\ifthenelse{\boolean{articletitles}}{\emph{See supplemental material at [link
  inserted by publisher] for the other projections, the slice plots, the fit
  parameters and fit fractions of fit results}, }{}\relax
\mciteBstWouldAddEndPuncttrue
\mciteSetBstMidEndSepPunct{\mcitedefaultmidpunct}
{\mcitedefaultendpunct}{\mcitedefaultseppunct}\relax
\EndOfBibitem
\bibitem{Gershon:2022xnn}
LHCb collaboration, T.~Gershon,
  \ifthenelse{\boolean{articletitles}}{\emph{{Exotic hadron naming
  convention}}, }{}\href{http://arxiv.org/abs/2206.15233}{{\normalfont\ttfamily
  arXiv:2206.15233}}\relax
\mciteBstWouldAddEndPuncttrue
\mciteSetBstMidEndSepPunct{\mcitedefaultmidpunct}
{\mcitedefaultendpunct}{\mcitedefaultseppunct}\relax
\EndOfBibitem
\end{mcitethebibliography}
\end{document}